\documentclass[showpacs,amsmath,amssymb,aps,twocolumn,longbibliography,pra]{revtex4-2}
\usepackage{graphicx}
\usepackage[colorlinks=true,citecolor=blue,linkcolor=magenta]{hyperref}
\usepackage[usenames]{color}
\usepackage{relsize}
\usepackage[english]{babel}
\usepackage{enumerate}
\usepackage{bbm}
\usepackage{xcolor}

\usepackage{framed}
\usepackage{bm}

\newcommand{\ket}[1]{\left| #1 \right>} 
\newcommand{\bra}[1]{\left< #1 \right|} 

\newcommand {\grsim} {\ {\raise-.5ex\hbox{$\buildrel>\over\sim$}}\ }
\newcommand {\lessim} {\ {\raise-.5ex\hbox{$\buildrel<\over\sim$}}\ }

\newcommand{\operator}{\scalebox{1.25}{$\delta$}}
\newcommand{\error}{\scalebox{1.5}{$\epsilon$}}
\usepackage{xcolor}
\usepackage{siunitx}
\usepackage{booktabs}

\newcommand{\nocontentsline}[3]{}
\newcommand{\tocless}[2]{\bgroup\let\addcontentsline=\nocontentsline#1{#2}\egroup}

\newcommand{\RN}[1]{%
  \textup{\uppercase\expandafter{\romannumeral#1}}%
}

\usepackage{xr}
\begin{document}
\title{A protocol to characterize errors in quantum simulation of many-body physics}
\author{Aditya Prakash$^{1,2}$}
\author{Bharath Hebbe Madhusudhana$^{3}$ }

\affiliation{$^{1}$National Institute of Science Education and Research Bhubaneswar, Jatni, Odisha 752050, India}
\affiliation{$^{2}$ Homi Bhabha National Institute, Training School Complex, Anushakti Nagar, Mumbai 400094, India}
\affiliation{$^{3}$MPA-Quantum, Los Alamos National Laboratory, Los Alamos, NM-87544, United States}


\begin{abstract} 
Quantum simulation of many-body systems, particularly using ultracold atoms and trapped ions, presents a unique form of quantum control --- it is a direct implementation of a multi-qubit gate generated by the Hamiltonian.  As a consequence, it also faces a unique challenge in terms of benchmarking, because the well-established gate benchmarking techniques are unsuitable for this form of quantum control. Here we show that the symmetries of the target many-body Hamiltonian can be used to benchmark and characterize experimental errors in the quantum simulation. We consider two forms of errors: (i) unitary errors arising out of systematic errors in the applied Hamiltonian and (ii) canonical non-Markovian errors arising out of random shot-to-shot fluctuations in the applied Hamiltonian. We show that the dynamics of the expectation value of the target Hamiltonian itself, which is ideally constant in time, can be used to characterize these errors. In the presence of errors, the expectation value of the target Hamiltonian shows a characteristic thermalization dynamics, when it satisfies the operator thermalization hypothesis (OTH). That is, an oscillation in the short time followed by relaxation to a steady-state value in the long time limit. We show that while the steady-state value can be used to characterize the coherent errors, the amplitude of the oscillations can be used to estimate the non-Markovian errors.  We develop scalable experimental protocols to characterize these errors. 
\end{abstract}

\maketitle

\textit{Introduction:} In quantum simulation, the properties of a  many-body Hamiltonian are simulated using a highly controllable quantum system. This includes simulating ground-state properties via adiabatic sweeps and time dynamics using a quench. Both of them have been demonstrated on various platforms including ultracold atoms in optical lattices~\cite{RevModPhys.80.885, Gross995}, ion traps~\cite{RevModPhys.93.025001, Zhang_2017, Blatt2012} and arrays of Rydberg atoms in optical tweezers~\cite{Ebadi_2021, Bernien_2017}. They produce a multi-qubit gate --- an entangling quantum gate acting on a large quantum system, usually the set of all qubits. They can expand the gate-set of quantum control in general and can be used to develop digital-analog quantum computers~\cite{PhysRevA.101.022305, garciadeandoin2023digitalanalog}. They can also be used to achieve practical quantum advantage by benchmarking new ansatz-based efficient classical algorithms to solve for the properties of the same Hamiltonians~\cite{PRXQuantum.2.040325}. An important limitation to all of these applications is the experimental errors incurred, which need to be benchmarked~\cite{daley2022practical, Hauke_2012}.

\begin{figure}[h!]
    \centering
    \includegraphics[scale=0.56]{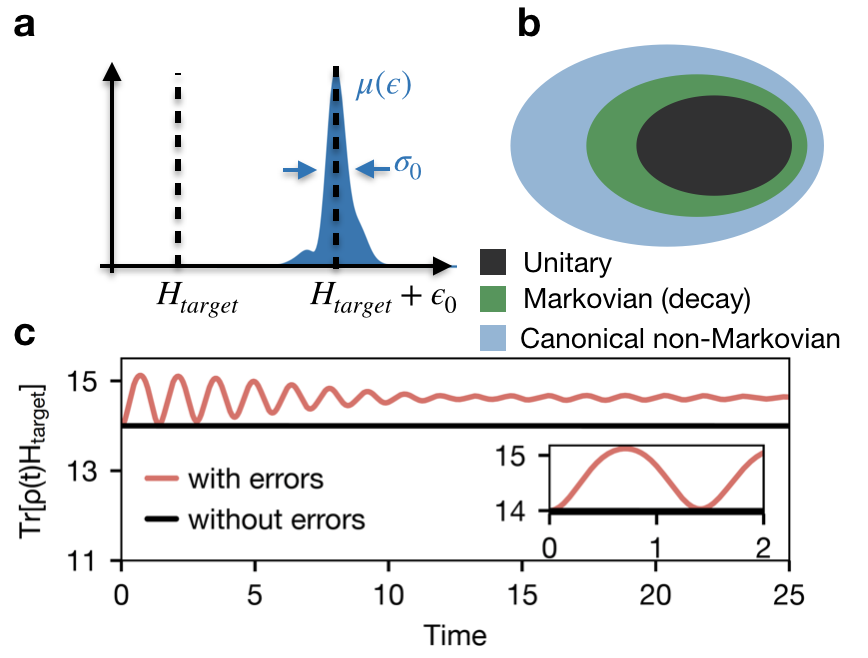}
    \caption{\textbf{Errors in quantum simulation:} \textbf{a.} Errors, $\error$ in the implementation of a target Hamiltonian $H_{target}$ are modelled by a distribution $\mu(\error)$ in the space of Hamiltonians. The mean $\error_0$  is the systematic part and the fluctuations $\sigma_0$ is the random part. \textbf{b.} $\error_0$ causes a unitary error (black). The shot-to-shot fluctuations cause canonical non-Markovian errors (blue). Any decay modelled by a Lindblad master equation causes Markovian errors (green). \textbf{c.} The dynamics of $H_{target}$ with and without errors. Without errors, it is trivially constant; with errors, it shows short-lived oscillations which relax the the long-time limit.  The steady-state value can be used to characterize the unitary errors and the oscillations can be used to estimate the non-Markovian errors.  }
    \label{Fig1}
\end{figure}

One and two-qubit gates can be benchmarked via a process tomography~\cite{Nielsen_2021}. At the device level, one can use randomized benchmarking and its variants~\cite{PhysRevLett.106.180504, PhysRevA.77.012307, PhysRevLett.123.030503, Proctor_2021}. However, neither of these are suited for benchmarking quantum simulation of a particular many-body Hamiltonian. Alternatives include cross-platform benchmarking and Hamiltonian learning~\cite{PhysRevLett.122.020504, PhysRevLett.124.160502, Wang_2017, Yu_2023}. However, the former requires an exponentially large number of measurements~\cite{PhysRevLett.124.010504, PRXQuantum.2.010102} and the latter assumes that the dynamics are unitary. 

Another alternative is to use random matrix theory.  A  state produced after a long time evolution under a typical many-body Hamiltonian has universal statistical properties, which maybe violated if there are experimental errors~\cite{Cotler_2023, Choi_2023, mark2023benchmarking, shaw2023benchmarking}. In this letter we develop a set of new protocols to characterize errors in quantum simulation of many-body systems. Our protocols are based on symmetries of the target Hamiltonian~\cite{madhusudhana2023benchmarking2} and are designed to estimate both the coherent (systematic) errors and non-Markovian errors (Fig.~\ref{Fig1}).  Moreover, they are scalable in the system size,  because, don't rely on a classical simulation. 

\textit{Problem Setup:} We consider the following general model of errors. Let $H_{target}$ be the target Hamiltonian, acting on $N$ qubits. The ideal unitary acting on a $2^N$ dimensional space generated by quench dynamics under this Hamiltonian would be $\Phi_{target}: \rho \mapsto e^{-i t H_{target}}\rho e^{it H_{target}}$, where $\rho$ is a mixed state of the $N$ qubits. However, due to errors, the real gate would be:

\begin{equation}
    \Phi_{expt}: \rho \mapsto \int d\mu(\error) \rho_{\epsilon}(t)
\end{equation}
Here, $\mu(\error)$ is a measure over the space of $N-$ qubit Hamiltonians and $\rho_{\epsilon}(t)$ is the solution to the equation
\begin{equation}\label{expt_gate}
    \dot{\rho}_{\epsilon} = -i [H_{target}+\error, \rho_{\epsilon}] + \sum_j L_j \rho_{\epsilon} L_j^{\dagger} -\frac{1}{2}\{\rho_{\epsilon}, L_j^{\dagger}L_j\}
\end{equation}
with $\rho_{\epsilon}(0)=\rho$.  $\error$ models the error in the applied Hamiltonian. We assume that this error can vary over samples, i.e., between experimental shots and therefore is modelled by a distribution $\mu(\error)$. The mean value, $\int \error d\mu(\error) = \error_0$ of this distribution, is the systematic error and the variance, $\int || \error-\error_0||^2 d\mu(\error) = \sigma_0^2$ represents the fluctuation, i.e., the random errors and it contributes to non-Markovian errors in the gate. Here, $||X||$ is an operator norm, the choice of which we discuss later. Note that at this point we donot make any assumption regarding the nature of the distribution $\mu(\error)$.


$L_i$ are Lindblad jump operators, representing decay processes which contribute to a Markovian error. Through the rest of this paper,  we assume that $L_i=0$, and focus on estimating the unitary and non-Markovian errors ($\error_0$ and $\sigma_0$). While this assumption is not experimentally justified, it makes the development of the theory simpler which can then be applied to real experiments after either minimizing or mitigating the Markovian errors. Moreover, the Markovian errors can be measured \textit{independently} and \textit{exclusively} using double stochasticity violation~\cite{madhusudhana2023benchmarking2}. Here, we focus on the problem of developing a protocol to estimate the first two moments of $\mu(\error)$, i.e., $\error_0$ and $\sigma_0$. We will show that the symmetries of the Hamiltonian $H_{target}$ can be used to develop such protocols.

\begin{figure}[h!]
    \includegraphics{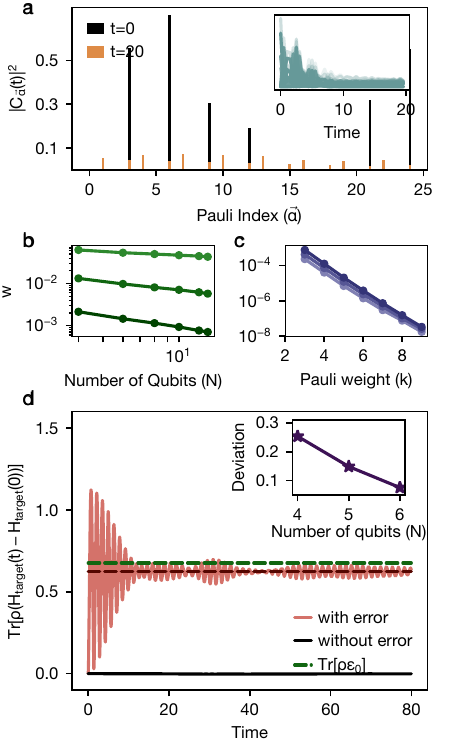}
    \caption{\textbf{Operator thermalization and typicality of commuting errors:} \textbf{a.} Coefficients $|C_{\vec{\alpha}}(t)|^2$ in $\error(t) = \sum_{\vec{\alpha}}C_{\vec{\alpha}}(t) P_{\vec{\alpha}}$ for $t=0$ (black bars) and long time ($t=20$, orange bars). $C_{\vec{\alpha}}(0)$ were chosen randomly and the time evolution was done with $N=8$ qubits, under the Hamiltonian in Eq.~(\ref{tweezer}). Inset shows $|C_{\vec{\alpha}}(t)|^2$ as a function of $t$, for $20$ values of $\vec{\alpha}$.  \textbf{b.} The weight of the commuting part, $w$ (see text, Eq.~(\ref{weight})) as a function of system size. The three lines correspond to $k(\vec{\alpha})=1, 2, 3$ \textbf{c.} $w$ for random $\vec{\alpha}$'s with different weights. The three lines correspond to $N=10, 11, 12$. \textbf{d.} Time evolution of $\text{Tr}[\rho (H_{target}(t)-H_{target}(0))]$ for the case with error (red curve) compared with the case without error (black line). The purple dashed line represents the thermal value and the green dashed line represents $\text{Tr}[\rho \error_0]$. The gap between these two represents the small $\error_{com}$.  The inset shows the gap between the thermal value and $\text{Tr}[\rho \error_0]$, as a function of system size.  See ref.~\cite{supplements} for more details of the data.  }
    \label{Fig2}
\end{figure}

We illustrate the setup with a simple example. Consider the standard Hamiltonian corresponding to $N$ Rydberg atoms in a tweezer array:
\begin{equation}\label{tweezer}
    H_{target} = \Omega \sum_j \sigma_{x,j} + \Delta \sum_j \sigma_{z,j} + \sum_{i}V\sigma_{z, i} \sigma_{z, i+1} 
\end{equation}
Here, $\sigma_{\alpha, j}= \mathbbm{1}^{\otimes j-1}\otimes \sigma_{\alpha} \otimes \mathbbm{1}^{\otimes (N-j)}$, with $\mathbbm{1}$ being the $2\times 2$ identity and $\alpha \in\{ x, y, z\}$.  This is a weight-1 pauli operator with a $\sigma_{\alpha}$ in the $j-$the position and identities on the rest. $\Omega$ and $\Delta$ are generated by the power and detuning of a coupling laser.  $V$ is produced by Rydberg blockade, approximated to nearest neighbours, by choosing the blockade radius.  See reef.~\cite{supplements} for more examples. The errors, for instance maybe caused by systematics and fluctuations of the laser power. In general, one can model the error as an operator within the span of a low-weight Pauli operator (see ref.~\cite{supplements} for physical examples). That is,
\begin{equation}\label{pauli_exp_eps}
    \error = \sum_{k(\vec{\alpha})\leq m} P_{\vec{\alpha}}\epsilon_{\vec{\alpha}}; \quad  \error_0 = \sum_{k(\vec{\alpha})\leq m} P_{\vec{\alpha}}\langle \epsilon_{\vec{\alpha}}\rangle_{\mu}
\end{equation}
Here, $\vec{\alpha}\in \{0, x, y, z\}^N$ and $P_{\vec{\alpha}}=\sigma_{\alpha_1}\otimes \cdots \otimes \sigma_{\alpha_N}$, with $\sigma_0=\mathbbm{1}$.  $k(\vec{\alpha})$ is the number of non-zero entries in $\vec{\alpha}$, i.e., the weight of $P_{\vec{\alpha}}$. Finally, $m\ll N$ is an integer that doesn't scale with $N$. Practically, one can assume $m\leq 4$. \\

%

\textit{Results:} Note that while $\Phi_{target}$ is a unitary map, $\Phi_{expt}$ is a general completely positive trace preserving (CPTP) map.  For reasons that will be clear soon, we switch to the Heisenberg picture. If an observable $\hat{O}$ commutes with the target Hamiltonian, i.e., $[H_{target}, \hat{O}]=0$, it follows that ideally $\hat{O}-\hat{O}(t)=0$. Here, $\hat{O}(t)$ is the time-evolved operator, defined by $\hat{O}(t)= \tilde{\Phi}_{target}(\hat{O})= e^{i t H_{target}}\hat{O} e^{-it H_{target}}$;  $\tilde{\Phi}_{target}$ is the map $\Phi_{target}$ in the Heisenberg picture.  We define $\tilde{\Phi}_{expt}$ as the map   $\Phi_{expt}$ the Heisenberg picture.  We choose $\hat{O}=H_{target}, H_{target}^2, \cdots$. When $H_{target}$ is a sum of low-weight Pauli operators, its expectation value can be measured efficiently in the lab.  Measuring $H_{target}^2, H_{target}^3, $ are less efficient. In this paper, we restrict to $H_{target}$. 

After time-evolution with errors, $H_{target}(t) =\tilde{\Phi}_{expt}(H_{target})$ is
\begin{equation}
    H_{target}(t) = \int d\mu(\error) e^{it (H_{target}+\error)}H_{target}e^{-it (H_{target}+\error)}
\end{equation}
Note that $e^{it (H_{target}+\error)}(H_{target} + \error) e^{-it (H_{target}+\error)} = (H_{target} + \error)$. Thus, it follows that
\begin{equation}\label{eq5}
    H_{target}(t) - H_{target}(0) = \int \error d\mu(\error) - \int \error(t) d\mu(\error)
\end{equation}
Here, $\error(t) =  e^{it (H_{target}+\error)} \error e^{-it (H_{target}+\error)}$. Eq.~(\ref{eq5}) simplifies to
\begin{equation}\label{the_eqn}
    H_{target}(t) - H_{target}(0) = \error_0 - \langle \error(t) \rangle_{\mu}
\end{equation}
where, $\langle \cdot \rangle_{\mu}$ is the averaging over $\mu(\error)$. For an initial state $\rho$,  an experimental measurement of $\text{Tr}[\rho ( H_{target}(t) - H_{target}(0))]$ gives us $\text{Tr}[\rho(\error_0- \langle \error(t)\rangle_{\mu})]$, which we show, can be used to characterize $\error_0$ and $\sigma_0$.\\

\begin{figure*}[]
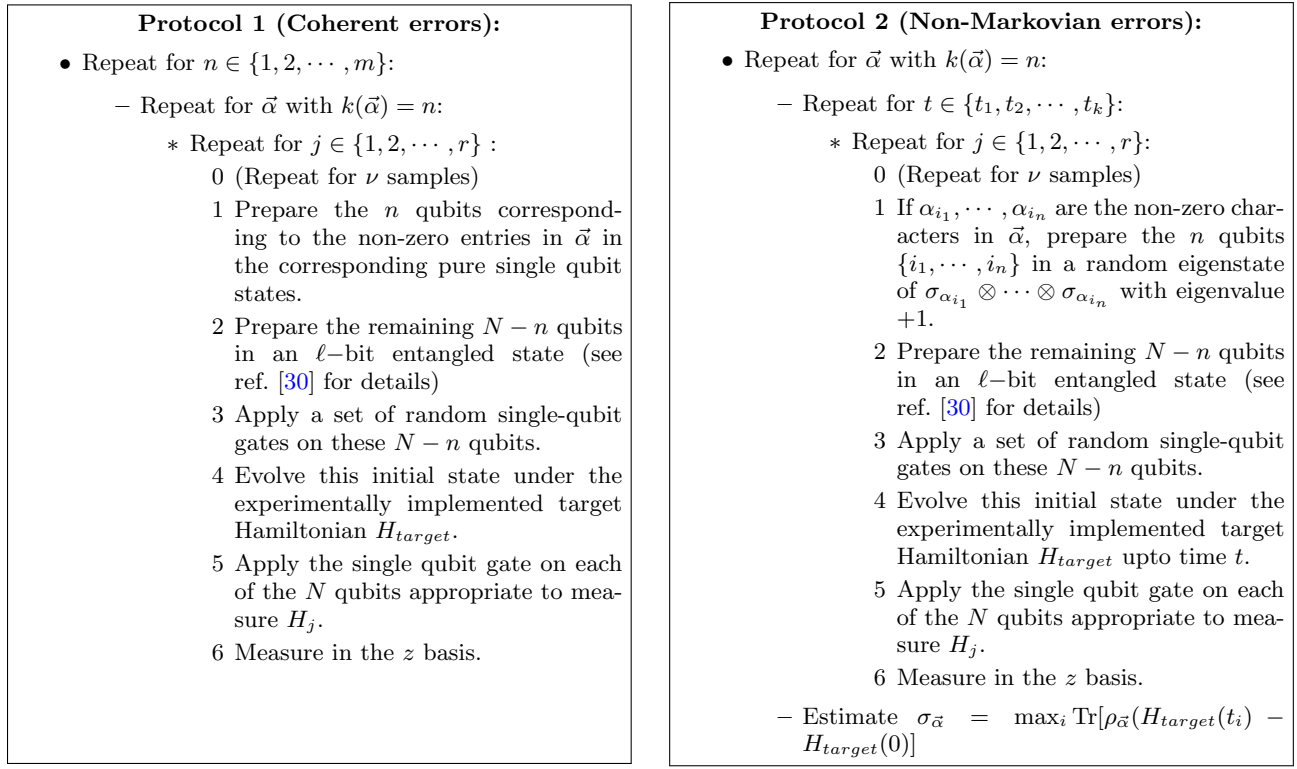

  \fbox{\begin{minipage}[l]{0.45\textwidth}
    \textbf{Protocol 1 (Coherent errors):}
    \begin{itemize}
\item Repeat for $n\in \{1, 2, \cdots, m\}$:
\begin{itemize}
    \item Repeat for $\vec{\alpha}$ with $k(\vec{\alpha})=n$:
    \begin{itemize}
        \item Repeat for  $j \in \{1, 2, \cdots, r\}$ :
        \begin{itemize}
            \item[0] (Repeat for $\nu$ samples)
            \item[1] Prepare the $n$ qubits corresponding to the non-zero entries in $\vec{\alpha}$ in the corresponding pure single qubit states.
            \item[2] Prepare the remaining $N-n$ qubits in an $\ell-$bit entangled state (see ref.~\cite{madhusudhana2023benchmarking1} for details)
            \item[3] Apply a set of random single-qubit gates on these $N-n$ qubits.
            \item[4] Evolve this initial state under the experimentally implemented target Hamiltonian $H_{target}$.
            \item[5] Apply the single qubit gate on each of the $N$ qubits appropriate to measure $H_j$. 
            \item[6] Measure in the $z$ basis.
        \end{itemize}
    \end{itemize}
\end{itemize}
\end{itemize}
\vspace{30pt}
  \end{minipage}}
  \hspace{8pt}
  \fbox{\begin{minipage}[l]{0.45\textwidth}
    \textbf{Protocol 2 (Non-Markovian errors):}
\begin{itemize}
    \item Repeat for $\vec{\alpha}$ with $k(\vec{\alpha})=n$:
    \begin{itemize}
        \item Repeat for $t\in \{t_1, t_2, \cdots, t_k\}$:
        \begin{itemize}
        \item Repeat for  $j \in \{1, 2, \cdots, r\}$: 
                \begin{itemize}
            \item[0] (Repeat for $\nu$ samples)
            \item[1] If $\alpha_{i_1}, \cdots, \alpha_{i_n}$ are the non-zero characters in $\vec{\alpha}$, prepare the $n$ qubits $\{i_1, \cdots, i_n\}$ in a random eigenstate of $\sigma_{\alpha_{i_1}}\otimes \cdots \otimes \sigma_{\alpha_{i_n}}$ with eigenvalue $+1$.
            \item[2] Prepare the remaining $N-n$ qubits in an $\ell-$bit entangled state (see ref.~\cite{madhusudhana2023benchmarking1} for details)
            \item[3] Apply a set of random single-qubit gates on these $N-n$ qubits.
            \item[4] Evolve this initial state under the experimentally implemented target Hamiltonian $H_{target}$ upto time $t$.
            \item[5] Apply the single qubit gate on each of the $N$ qubits appropriate to measure $H_j$. 
            \item[6] Measure in the $z$ basis.
        \end{itemize}
        
        \end{itemize}
        
        \item Estimate $\sigma_{\vec{\alpha}} = \max_{i} \text{Tr}[\rho_{\vec{\alpha}}(H_{target}(t_i)-H_{target}(0)]$

    \end{itemize}
\end{itemize}
  \end{minipage}}
  \caption{\textbf{Protocols:} Left panel: a  protocol  to extract the individual error coefficients $\langle \epsilon_{\vec{\alpha}}\rangle_{\mu}$, for low weight indices $\vec{\alpha}$.  Right panel: protocol to estimate  the magnitude of the fluctuations within a Pauli weight of $n$, $\sigma_n = \sqrt{\sum_{k(\vec{\alpha})=n} \sigma_{\vec{\alpha}}^2}$ and $\sigma_0^2 =\sum_n \sigma_n^2$.  }\label{protocols}
\end{figure*}

\textit{Characterizing $\error_0$:} This involves measuring the average values $\langle \epsilon_{\vec{\alpha}}\rangle_{\mu}$ (see Eq.~(\ref{pauli_exp_eps})).  Corresponding to the pauli operator $P_{\vec{\alpha}}$, let us define a state $\rho_{\vec{\alpha}} = \frac{1}{2^N}(\mathbbm{1}^{\otimes N} + P_{\vec{\alpha}})$.  For example, if $P_{\vec{\alpha}}=\sigma_x\otimes \mathbbm{1}^{\otimes N-1}$,  in  the state $\rho_{\vec{\alpha}}$, the first qubit is in the $|+\rangle$ state and the rest are maximally mixed. Note that $\langle \epsilon_{\vec{\alpha}}\rangle_{\mu} = \text{Tr}[\rho_{\vec{\alpha}} \error_0]$. Thus, if we measure $H_{target}(t)$ with the initial state $\rho_{\vec{\alpha}}$,it follows from Eq.~(\ref{the_eqn}) that,
\begin{equation}\label{basic_error_eq}
    \text{Tr}[\rho_{\vec{\alpha}}(H_{target}(t)-H_{target}(0))] = \langle \epsilon_{\vec{\alpha}}\rangle_{\mu} - \text{Tr}[\rho_{\vec{\alpha}} \langle \error(t)\rangle_{\mu}]
\end{equation}
In the following, we will show that under ergodic Hamiltonians, $\text{Tr}[\rho_{\vec{\alpha}} \langle \error(t)\rangle_{\mu}]\approx 0$ in the limit of large $t$. Thus, $\langle \epsilon_{\vec{\alpha}}\rangle_{\mu} \approx \text{Tr}[\rho_{\vec{\alpha}}(H_{target}(t)-H_{target}(0))]$ and it can be used to measure the error. The time evolution $\error(t)$ under the Hamiltonian $H_{target}+\error$ can be understood using \textit{operator thermalization}. We can write $\error$ as the sum of a commuting and a non-commuting part with $H_{target}+\error$. That is,
\begin{equation}
    \error = \error_{com} + \error_{noncom}
\end{equation}
where, $[H_{target}+\error, \error_{com}]=0$. 
$\error_{com}$ is uniquely defined as the diagonal blocks of $\error$ in the eigenbasis of $H_{target}+\error$. See ref.~\cite{supplements} for a rigorous and other equivalent definitions of $\error_{com}$. 


Under the time-evolution, $\error(t) = \error_{com} + \error_{noncom}(t)$, i.e., the commuting part does not evolve in time. The $mn-$th element of $\error_{noncom}(t)$ in the eigenbasis $\{\ket{E_n}\}$ of $H_{target}+\error$ is, $\bra{E_n}\error_{noncom}(t)\ket{E_m} = e^{-i(E_n-E_m)t}\bra{E_n}\error_{noncom}(0)\ket{E_m}$, the phases of which, in general can be assumed to be nearly random at large $t$. Thus,  in the Pauli basis, $\error_{noncom}(t) = \sum P_{\vec{\alpha}}C_{\vec{\alpha}}(t)$, the coefficients $C_{\vec{\alpha}}(t)=\frac{1}{2^N}\text{Tr}[P_{\vec{\alpha}} \error_{noncom}(t)]$ would be exponentially suppressed (Fig.~\ref{Fig2}a). Moreover, the averaging over $\mu(\error)$ further suppresses $\error_{noncom}(t)$. Fig~\ref{Fig2}a includes this additional suppression. Therefore,  after a sufficiently long time, $\text{Tr}[\rho_{\vec{\alpha}}(H_{target}(t)-H_{target}(0))] \approx \frac{1}{2^N}\text{Tr}[(\error_0-\error_{com})P_{\vec{\alpha}}]$ and therefore a measurement of the former can be used to characterize $\error_0-\error_{com}$.

An averaging over time can be used to further suppress the time-dependent part $\error_{noncom}(t)$. It is straightforward to see that 
$$
\lim_{T\rightarrow \infty} \frac{1}{T}\int_0^T dt\ \error_{noncom}(t) =0
$$
However, the rate of convergence may vary and has been studied extensively in the theory of operator thermalization. See ref.~\cite{supplements} for a brief discussion.  

In general, one cannot eliminate the commuting part $\error_{com}$ from the measured projection. However, one can show that the commuting part is vanishingly small for experimentally relevant cases. Let us define the magnitude of the commuting part, 
$$
w= \frac{|| \error_{com}||^2}{||\error||^2} = \frac{\sum_n |\bra{E_n}\error\ket{E_n}|^2}{||\error||^2}  
$$
If $\error =\epsilon_{\vec{\alpha_0}}P_{\vec{\alpha_0}}$, i.e., a particular Pauli, the weight is
\begin{equation}\label{weight}
    w= \frac{\sum_n |\bra{E_n}P_{\vec{\alpha_0}}\ket{E_n}|^2}{2^N} 
\end{equation}
Note that while this is the ratio of the magnitude squared of the projection of $\error$ onto the null space of the adjoint operator, $\text{ad}_{H_{target}+\epsilon}$($\text{ad}_A(X)=[A, X]$).  It can also be understood as the ratio of the projections on to the state $\rho_{\vec{\alpha}}$, i.e.,  $w = \text{Tr}[\rho_{\vec{\alpha}_0}\error_{com} ]/\text{Tr}[\rho_{\vec{\alpha}_0}\error ]$. If $\ket{E_n}$ are highly entangled, one can show that $w$ is vanishingly small.  To see this, let us consider  a random matrix product state (MPS) with bond dimension $\chi$, $\ket{\psi}$.  It follows from the typicality result for random MPS with a given bond dimension $\chi$ from refs~\cite{PhysRevA.81.032336, PhysRevA.82.052312} that
$$
\text{Pr}[|\bra{\psi}P_{\vec{\alpha}}\ket{\psi}|\geq \delta ]\leq c_1 e^{-c_2 \delta^2\frac{\chi}{N^2}}
$$
Therefore, if the bond dimension of the bulk eigenstates $\ket{E_n}$ grows faster than $O(N^2)$, the typical value of $|\bra{\psi}P_{\vec{\alpha}}\ket{\psi}|$ is suppressed to $O\left(\frac{N}{\sqrt{\chi}}\right)$ and so does $w$. Fig.~\ref{Fig2}b shows $w$ as a function of $N$ for $k(\vec{\alpha})=1,2,3$ for the Hamiltonian in Eq.~(\ref{tweezer}).

The data appears to converge polynomially down to zero. Fig.~\ref{Fig2}c shows $w$ for the same Hamiltonian, for various weights $k(\vec{\alpha})$, which appears to converge exponentially. This observation is also related to the features of operator thermalization. In fact, one can show that if the eigenstates of $H_{target}+\error$ have an entanglement depth of $\ell$, i.e., they are $\ell-$bit entangled states, then the magnitude of the commuting part $w$ of a Pauli operator with weight $k$ scales as $w\sim \frac{1}{(2^{\ell}+1)^k}$ (see ~\cite{supplements} for a proof and details). Fig.~\ref{Fig2}d shows a comparison between $\text{Tr}[\rho_{\vec{\alpha}}(H_{target}(t)-H_{target}(0)]$, its thermal value $\frac{1}{T}\int_0^T dt \text{Tr}[\rho_{\vec{\alpha}}(H_{target}(t)-H_{target}(0)]$ and the error coefficient $\langle \epsilon_{\vec{\alpha}}\rangle_{\mu}$. We use these results to construct a protocol to estimate $\langle \epsilon_{\vec{\alpha}}\rangle_{\mu}$.  See Fig.~\ref{protocols}, left panel.

\begin{figure}
    \centering
    \includegraphics[scale=1]{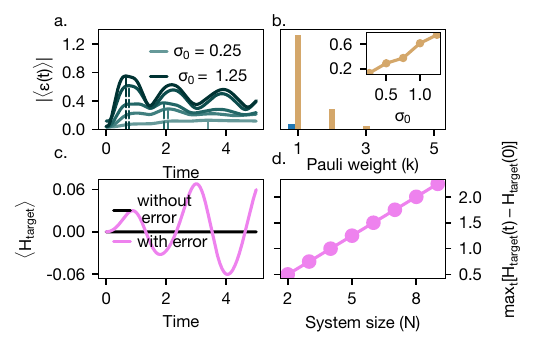}
    \caption{\textbf{Numerical experiment:} \textbf{a} Time evolution of $||\error(t)||$, with five values of $\sigma_{0}$ ranging from $0.25$ to $1.25$. \textbf{b} Distribution of $\max_t||\error(t)||$ across various pauli weights (orange).  The blue bar shows the same data at $t=0$. The inset presents value of maximum of $||\error(t)||$ shown in a versus $\sigma_{0}$. \textbf{c} Time evolution of $H_{target}$ in the short time, with $\error_0=0$.  \textbf{d} The maximum value over time of $H_{target}$ as a function of system size $N$ for a particular error $\error = \epsilon \sum_j \sigma_{z, j}$, without averaging.  The linearity comes from $|\error(t)||$ being linear in $N$. }  
    \label{Fig3}
\end{figure}


\textit{Estimating $\sigma_0$:} We now show that the information of $\sigma_0$ can be extracted from short-time dynamics after correcting for the systematics $\error_0$.  We will show that, for $\error_0=0$, 
\begin{equation}\label{ineq}
    O(\sigma_0) \leq  \max_t || \langle \error(t) \rangle_{\mu} || \leq  \sigma_0
\end{equation}
To show the upper bound, note that the time evolution of a particular $\error$ is unitary and therefore, $||\error(t)||=||\error||$. This, along with the triangle inequality, gives 
$$
|| \langle \error(t) \rangle_{\mu} || \leq \int d\mu(\error) ||\error || \leq  \sigma_0
$$
The last inequality follows from the identity $\int |X| \leq \sqrt{\int |X|^2}$. The lower bound is more nontrivial to show. While similar lower bounds can be shown easily for scalar functions, owing to the nature of non-commutative analysis, showing lower bounds on matrix functions can take us in to deep waters. However, it turns out that we can show such lower bounds for $\error(t)$ using typical statistics for the level gaps for $H_{target}$. The below theorem shows a lower bound.

\noindent\textbf{Theorem 1: }  If the measure $\mu$ is \textit{light-tailed}, there exist constants $a, b>0$, \textit{independent} of $N$ such that at $t_0 = \frac{a}{\sigma_0}$, 
\begin{equation}
    ||\langle \error (t_0)\rangle_{\mu} ||\geq b \sigma_0
\end{equation}

The $N-$independence of $b$ is crucial for scalabiliy and it is related to the typical value of the differences $E_i-E_j$ between energies of $H_{target}$~\cite{santos2017nonequilibrium, haake2010quantum}. While the proof of this theorem  is interesting in its own right, it is also very technical and therefore we reluctantly place it in the supplementary information~\cite{supplements}.

We show this result also using a numerical experiment.  In Fig.~\ref{Fig3}a, we show the time evolution of $ ||\error(t)||$, under the Schatten 2-norm for $\error = \epsilon \sigma_{\alpha_0}$ and gaussian random variable $\epsilon$.  Note that the time at which the maximum is reached decreases with increasing $\sigma_0$ (dashed lines). In Fig.~\ref{Fig3}b inset, we show the value of the maximum as a function of $\sigma_0$. The linearity is apparent with $\max_t ||\error(t)|| \approx 0.6 \sigma_0$.  Note that the evolution under $H_{target}$ is expected to scramble the error $||\error(t)|| $ across weights, making it nontrivial to measure. However, as we showed earlier, the maximum is reaches at time $\sim 1/\sigma_0$ and if this time is smaller than the thermalization timescale of $H_{target}$, one can efficiently estimate  $\max_t ||\error(t)||$. In this case, a measurement of $\text{Tr}[\rho (H_{target}(t)-H_{target}(0))]$ can be used to estimate $\sigma_0$, by using $\rho=\rho_{\vec{\alpha}}$ for low-weight $\vec{\alpha}$. See Fig.~\ref{protocols} for a detailed protocol.  In Fig.~\ref{Fig3}c,d we run a similar numerical experiment with more general error,  involving multiple Paulis. We find that the deviation, $H_{target}(t)-H_{target}(0)$ is infact linear in $\sigma_0$ even in this case.   \\



\textit{Conclusions:} We have developed a scalable benchmarking technique for quantum simulators using measurements of the target Hamiltonian. It is effective against systematic errors in the Hamiltonian and non-Markovian errors generated by shot-to-shot fluctuations.  This scheme can be implemented on Rydberg atoms trapped in a tweezer array or on optical lattices loaded with neutral atoms.  We emphasize that our protocols don't need a classical simulation and are therefore scalable and can be used in regimes inaccessible to classical computation. This is crucial for practical quantum advantage~\cite{PRXQuantum.2.040325}.  A major experimental challenge is to characterize and/or mitigate the Markovian errors,  which will interfere with the characterization of the errors presented in this paper. The timescales of Markovian  errors observed have been long enough to observe even many-body localization dynamics~\cite{scherg2021observing}. They can be characterized independently using double-stochasticity violation~\cite{madhusudhana2023benchmarking2}. Moreover, quantum error mitigation techniques have been developed recently, specifically targeting the Markovian errors generated by coupling to environment~\cite{donvil2023quantum, PhysRevApplied.15.034026}. 

Another important future direction is to extend the theory for time-dependent errors. As seen in ref~\cite{PhysRevLett.130.010201}, experimental errors are often time dependent.  Protocols for a more detailed characterization of the covariance matrix of $\error$ would also be useful in order to optimize errors in quantum control.

\section*{Acknowledgements}
We thank Malcolm Boshier, Karatzyna Krzyzanowska and Monika Aidelsburger for fruitful discussions.  The research presented in this article was supported by the Laboratory Directed Research and Development program of Los Alamos National Laboratory under project number 20230779PRD1.  A.P. wishes to express his gratitude to his parents for their constant support.

\paragraph*{\textbf{Competing interests}} The authors declare no competing interests. 

\bibliography{References}

\appendix

\cleardoublepage

\setcounter{figure}{0}
\setcounter{page}{1}
\setcounter{equation}{0}
\setcounter{section}{0}

\renewcommand{\thepage}{S\arabic{page}}
\renewcommand{\thesection}{S\arabic{section}}
\renewcommand{\theequation}{S\arabic{equation}}
\renewcommand{\thefigure}{S\arabic{figure}}
\onecolumngrid
\begin{center}
\huge{Supplementary Information}
\vspace{5mm}
\end{center}
\twocolumngrid
\normalsize
\tableofcontents

\section{Examples}

\textit{Example 1: Rydberg atoms} Consider a linear array of Rydberg atoms in a tweezer array. One can simulate a transverse field Ising model by using a dressing laser and a Rydberg laser. The target Hamiltonian would be
\begin{equation}\label{tweezer}
    H_{target} = \Omega \sum_j \sigma_{x,j} + \Delta \sum_j \sigma_{z,j} + \sum_{i}V\sigma_{z, i} \sigma_{z, i+1} 
\end{equation}
Here, $\sigma_{\alpha, j}= \mathbbm{1}^{\otimes j-1}\otimes \sigma_{\alpha} \otimes \mathbbm{1}^{\otimes (N-j)}$, with $\mathbbm{1}$ being the $2\times 2$ identity and $\alpha \in\{ x, y, z\}$.  This is a weight-1 pauli operator with a $\sigma_{\alpha}$ in the $j-$the position and identities on the rest. $\Omega$ and $\Delta$ are generated by the power and detuning of a coupling laser.  $V$ is produced by Rydberg blockade, approximated to nearest neighbours, by choosing the blockade radius. In the implementation of this model, the most common forms of errors can be modelled as 
\begin{equation}
\begin{split}
    \error &=  \sum_j \delta\Omega_j \sigma_{x,j} +  \sum_j\delta\Delta_j \sigma_{z,j}\\
    &+ \sum_{i, j}\delta V_{ij}\sigma_{z, i} \sigma_{z, j} + \sum_{ijk}\delta V_{i,j,k} \sigma_{z, i} \sigma_{z, j} \sigma_{z, k}+\cdots 
\end{split}
\end{equation}
Here, $\delta\Omega_j$ and $\delta\Delta_j$ are random numbers describing the spatial and shot-to-shot variation of the laser power and frequency. $\delta V_{ij}$, $\delta V_{ijk}$ etc. represent the residual Rydberg blockade and any possible multi-qubit interactions, although the latter can be assumed to be restricted to a $3-$qubit terms. They all may be modelled as Gaussian random variables, in which case, $\mu(\error)$ would be a multi-variable Gaussian distribution. If the mean values over several shots of any of these random variables is non-zero, then we will have a systematic error, $\error_0$
\begin{equation}
\begin{split}
    \error_0 &= \int \error d\mu(\error) =  \sum_j \langle \delta\Omega_j \rangle_{\mu} \sigma_{x,j} +  \sum_j \langle \delta\Delta_j \rangle_{\mu} \sigma_{z,j}\\
    &+ \sum_{i, j}\langle \delta V_{ij}\rangle_{\mu} \sigma_{z, i} \sigma_{z, j} + \sum_{ijk}\langle \delta V_{i,j,k}\rangle_{\mu} \sigma_{z, i} \sigma_{z, j} \sigma_{z, k}+\cdots 
\end{split}
\end{equation}
Here, $\langle \cdot \rangle_{\mu}$ represents the average of these random variables over the model distribution.


\textit{Example 2: Bosons in a lattice: } Consider a lattice with $N$ sites loaded with bosonic atoms, typically realized with ultracold atoms in an optical lattice. This system can be used to simulate the Bose-Hubbard model and its variants. The Hamiltonian is
\begin{equation}
\begin{split}
    H_{target} &= -J\sum_j \hat{a}_j^{\dagger}\hat{a}_{j+1} + \hat{a}_{j+1}^{\dagger}\hat{a}_{j} \\
    &+ \sum_j \Delta_j \hat{a}_j^{\dagger}\hat{a}_{j} +\frac{U}{2}\sum_j \hat{a}_j^{\dagger}\hat{a}_{j}(\hat{a}_j^{\dagger}\hat{a}_{j}-1)
\end{split}
\end{equation}
Here, $\hat{a}^{\dagger}_j$ is the creation operator for site $j$. $J$ is the hopping rate, controlled by the lattice depth.  $\Delta_j$ is an on-site potential. When $U\rightarrow \infty$, one can re-write this Hamiltonian in terms of spin-$1/2$ operators:
\begin{equation}
    H_{target} = -\frac{J}{2} \sum_j (\sigma_{x, j}\sigma_{x, j+1}+\sigma_{y, j}\sigma_{y, j+1} ) +\sum_j \Delta_j \sigma_{z, j}  
\end{equation}
Here, $\hat{a}_j^{\dagger}$ is mapped to $\sigma_{+, j}$. The standard errors can be modelled as
\begin{equation}
    \error =   \sum_{ij} \delta J_{ij} (\sigma_{x, i}\sigma_{x, j}+\sigma_{y, i}\sigma_{y, j} ) +\sum_j \delta \Delta_j \sigma_{z, j}  
\end{equation}
Here, $\delta J_{ij}$ are random numbers representing errors in the nearest neighbour, next nearest neighbour, next-next nearest neighbour hopping, etc. Note that in most cases, hopping processes with a range longer than the next-nearest neighbour are quite insignificant. $\delta \Delta_j$ are random numbers representing on-site potentials due to laser power fluctuations, etc. It is important to note that having double occupancies due to a finite Hubbard interaction $U$ can be a significant source of error. Under the mapping to spin-$1/2$ system, this will appear as a non-Markovian error due to leakage to other levels. However, one can model these errors effectively by mapping  each site to a qudit, instead of a qubit. That is, $\hat{a}_j^{\dagger}$ can be mapped to $S_+$, where the latter is the ladder operator for a $d-$level system, defined with appropriate coefficients. While they are beyond the scope of the main text, the results  can be generalized to qudits as well.

\section{Definitions of the commuting part}\label{commuting_part}
In this section, we will provide three rigorous definitions of the commuting part $A_{com}$ of an operator $A$ with respect to a given operator $M$ and show their equivalence to each other.

\noindent \textit{Definition 1:} If $\{\Pi_n\}$ are the eigen projectors of $M$, 
$$
A_{com} = \sum_n \Pi_n A\Pi_n 
$$

\noindent \textit{Definition 2:} $A_{com}$ is the operator that minimizes the function $f(X) = ||X-A||^2$ over all operators  $X$ that commute with $M$. 

\noindent \textit{Definition 3:} $A_{com}$ is the projection of $A$ onto the null space of the superoperator $\text{ad}_{M}$.

We begin by showing the equivalence of definitions 1 and 2. We write $A$ as
$$
A = \sum_n \Pi_n A\Pi_n + Q
$$
where $Q=A-\sum_n \Pi_n A\Pi_n$. Hereafter, we refer to the operator $\sum_n \Pi_n A\Pi_n$ as $P$. This is indeed the $A_{com}$ according to definition 1. The goal is to show that $P$ is the commuting part of $A$ with $M$, according to definition 2 as well. We begin by showing that $P$ and $Q$ are orthogonal. Indeed,
\begin{equation}
   \begin{split}
       \text{Tr}[PQ] =& \text{Tr}[(\sum_n \Pi_n A\Pi_n)(A-\sum_m\Pi_m A\Pi_m)] \\
       &=\sum_n\text{tr}[\Pi_n A \Pi_n A] -\sum_{m, n}\text{Tr}[\Pi_n A\Pi_n\Pi_m A\Pi_m]
   \end{split} 
\end{equation}

\noindent Using the orthogonality of the projectors, $\Pi_n \Pi_m=0$ unless $n=m$. Therefore, 
$$
\text{Tr}[PQ] =  \sum_n\text{tr}[\Pi_n A \Pi_n A] -\sum_{ n}\text{Tr}[\Pi_n A\Pi_n^2 A\Pi_n]
$$
We now use the fact that $\Pi_n^2=\Pi_n$ and that $\text{Tr}[\Pi_n A\Pi_n^2 A\Pi_n] = \text{Tr}[\Pi_n\Pi_n A\Pi_n^2 A]$, to show that the above is zero. If $X$ is an operator that commutes with $M$, it follows that 
$$
X= \sum_n \Pi_n X\Pi_n
$$
Therefore, 
\begin{equation}
\begin{split}
    ||X-A||^2 & = \text{Tr}[(X-A)(X-A)] = \text{Tr}[A^2] + \text{Tr}[X^2] - 2\text{Tr}[XA]\\
    & = \text{Tr}[A^2] + \sum_n [X^2] - 2\sum_n \text{Tr}[\Pi_n X\Pi_nA]
\end{split}
\end{equation}
We can now use the cyclic permutation rule of commutators:
\begin{equation}
    \begin{split}
    ||X-A||^2 & = \text{Tr}[A^2] + \text{Tr}[X^2] - 2\sum_n \text{Tr}[X\Pi_nA\Pi_n]\\
    & = \text{Tr}[A^2] +\text{Tr}[ X^2] - 2\text{Tr}[XP]\\
    & = \text{Tr}[A^2]-\text{Tr}[P^2] + ||P-X||^2
\end{split}
\end{equation}
Note that $P$ commutes with $M$ and therefore, $X=P$ is a legitimate choice. This function is quadratic and is minimized when $X=P$. Thus, definition 2 gives the same result as definition 1. 

Finally, we will show that definition 3 is equivalent to definition 2. The adjoint map $\text{ad}_M$ is defined as
$$
\text{ad}_M(X) =[M, X]
$$
Therefore, the null space of the superoperator $\text{ad}_M$ is the space of all operators that commute with $M$. Therefore, the projection of $A$ onto this space is indeed the operator from this space that minimizes $||A-X||^2$, which is the commuting part according to definition 2. 

\subsection{Algebra of the commuting part}
In this section, we discuss some of the basic algebraic properties of the commuting part of an operator w.r.t. a given operator. 

\noindent \textbf{Lemma 1:} If $M$ is a given operator, for operators $A$, $B$, it follows that
\begin{itemize}
    \item [i.] $(A+B)_{com} = A_{com} + B_{com}$
    \item[ii.] $(AB)_{com}=A_{com}B_{com} + (A_{noncom}B_{noncom})_{com}$
\end{itemize}
Where the commuting part is defined w.r.t. $M$.

\noindent \textbf{Proof:} We use the first definition of the commuting part. If $\Pi_n$ are the eigen projectors of $M$, 
\begin{equation}
    \begin{split}
        (A+B)_{com} &= \sum_n \Pi_n (A+B)\Pi_n \\
        &= \sum_n \Pi_n A\Pi_n +\Pi_n B\Pi_n = A_{com}+B_{com}
    \end{split}
\end{equation}

Moreover, note that
\begin{equation}
    \begin{split}
        A_{noncom} &= A-\sum_n \Pi_n A\Pi_n = \sum_n \Pi_n A- \sum_n \Pi_n A\Pi_n\\
        &=\sum_n \Pi_n A(1-\Pi_n)
    \end{split}
\end{equation}

\begin{equation}
    \begin{split}
        A_{com}B_{com} &= \sum_{n, m} \Pi_n A\Pi_n  \Pi_m B\Pi_m\\
        &=  \sum_n  \Pi_n A\Pi_n   B\Pi_n\\ 
         & = (AB)_{com} - \sum_n  \Pi_n A(1-\Pi_n) B\Pi_n \\
         &=(AB)_{com} - \sum_n  \Pi_n A(1-\Pi_n) (1-\Pi_n) B\Pi_n \\
         & = (AB)_{com} - \sum_n  \Pi_n (A_{noncom}B_{noncom})\Pi_n\\
         & = (AB)_{com}-(A_{noncom}B_{noncom})_{com}
    \end{split}
\end{equation}

The last equation follows from the fact that $\Pi_n \Pi_m=0$ if $n\neq m$.

\section{Operator thermalization}\label{op_therm}
In this section, we present a brief overview of the theory of operator thermalization. The investigation into how isolated, unitary quantum systems can appear to attain thermal equilibrium has long been a central concern within the field of statistical mechanics, with one of its cornerstones being the eigenstate thermalization hypothesis(ETH) \cite{ETH1,ETH2,ETH3,ETH4,ETH5}. However, considering the Heisenberg picture of evolution, recently, interest has shifted from states to operators, thus introducing the Operator Thermalization Hypothesis (OTH), which was first prominently mentioned in the ref. \cite{OTH1}. A meaningful comparison between ETH and OTH is presented in \cite{OTH_comparison}, where the authors use the transverse field Ising model to demonstrate examples exhibiting similarities and differences between ETH and OTH. The authors draw a parallel between ETH and OTH, elucidating that ETH leads to thermalization due to a chaotic spectrum allowing extensive Hilbert space exploration even with a simple operator, while OTH, as its counterpart, features an organized spectrum, requiring a complex operator for similar exploration.  For further investigation, the reader is encouraged to check \cite{OTH1, OTH_growth,craps2019energy}.

Let $\hat{O}$ be any operator acting on a quantum system, which evolves according to Hamiltonian $H$. Under the Heisenberg picture, the operator evolves in time as $\hat{O}(t) = e^{i H t}\hat{O}e^{-i H t}$. To understand this time evolution, let us write the operator as a sum of the commuting and the non-commuting parts w.r.t. $H$, as defined above: $\hat{O} = \hat{O}_{com}+\hat{O}_{noncom}$. The commuting part is time-invariant. Therefore, 
$$
\hat{O}(t) =  \hat{O}_{com}+\hat{O}_{noncom}(t) =  \hat{O}_{com}+e^{i H t}\hat{O}_{noncom}e^{-i H t}
$$
We now show that 
$$
\lim_{T\rightarrow \infty} \frac{1}{T}\int_0^T\hat{O}(t) = \hat{O}_{com}
$$
That is, the non-commuting part of the operator vanishes in the thermal limit. To see this, note that the $ij$-th element, in the eigenbasis of $H_{target}$, of $\hat{O}_{noncom}(t)$ is
$$
\bra{E_i}\hat{O}_{noncom}(t)\ket{E_j} = e^{i ( E_i - E_j )t}\bra{E_i}\hat{O}_{noncom}\ket{E_j}, \ \ E_i \neq  E_j
$$
Note that when $E_i = E_j$, $\bra{E_i}\hat{O}_{noncom}\ket{E_j}=0$. It now follows that the time average of the non-commuting part vanishes in the limit of large $T$. 

The question of the rate and nature of this convergence is a more complex one, for it depends on the distribution of the gaps, $ E_i-E_j$. These gaps are infact the eigenvalues of $\text{ad}_H$. The existence of a large number of very small gaps implies very long-time non-trivial dynamics and slow convergence to the thermal value. 

\section{Typicality of the commuting part}
In this section, we discuss the connection between thermalization and the typical values of the weight of the commuting part. Let $P_{\vec{\alpha}}$ be a pauli operator and $H$ be a many-body Hamiltonian. If $\ket{E_n}$ are the eigenstates of $H$, the magnitude $w$ of the commuting part of $P_{\vec{\alpha}}$ w.r.t $H$ is given by
$$
w_{\vec{\alpha}} =\frac{1}{2^N}\sum_n (\text{Tr}[P_{\vec{\alpha}}\ket{E_n}\bra{ E_n}])^2
$$
This can be understood as the second moment of the coefficients of $P_{\vec{\alpha}}$ in the density matrices $\ket{ E_n}\bra{E_n}$. That is, if we write
$$
\ket{ E_n}\bra{E_n} = \sum_{\vec{\beta}} P_{\vec{\beta}} C_{\vec{\beta}, n}
$$
the magnitude of the commuting part is
$$
w_{\vec{\alpha}} = 4^N \langle C^2_{\vec{\alpha}, n}\rangle_{\{n\}}
$$
Here, the average is over the eigenstates of $H$. We can now show some typicality results, if we assume that the eigenstates of $H$ form a $2-$design corresponding to some underlying distribution of states. For instance, we can consider random states with a fixed entanglement. We first begin with the simple case of Haar random states. The second moment of $C_{\vec{\alpha}}$ over Haar random states is straightforward to compute. Under Haar randomness, $\langle C_{\vec{\alpha}}^2\rangle$ must be independent of $\vec{\alpha}$. Moreover, $C_{\vec{0}}=1/2^N$ and 
$$
\sum_{\vec{\alpha}} C_{\vec{\alpha}}^2 =\frac{1}{2^N}
$$
Combining the two, we obtain:
$$
\langle C_{\vec{\alpha}}^2\rangle = \frac{1}{2^N(4^N-1)}\left(1-\frac{1}{2^N}\right) =\frac{1}{4^N(2^N+1)}
$$

\begin{figure}
    \centering
    \includegraphics{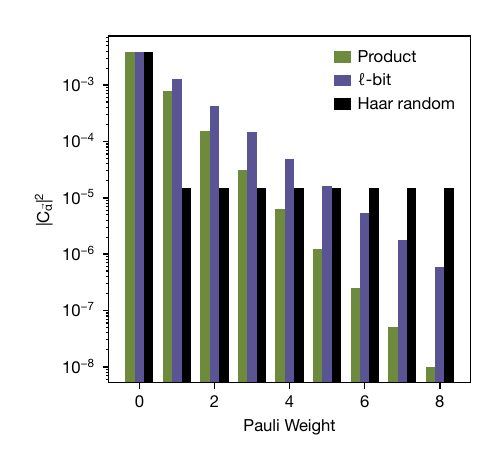}
    \caption{\textbf{Entanglement and Pauli coefficients:} The average of $|C_{\vec{\alpha}}|^2$ over various ensembles. For Haar random states, it takes a uniform value, $\frac{1}{4^N(2^N+1)}$. For $\ell-$bit states, it falls of exponentially $\frac{1}{4^N(2^{\ell}+1)^k}$ (see text).}
    \label{FigS2}
\end{figure}
However, the eigenstates of most Hamiltonians are not $2-$designs of Haar random states. They are more likely to have a fixed entanglement, but not highly entangled unlike Haar random states. Let us consider the case of products of $\ell-$bit Haar random states, i.e., a tensor product of $m =N/\ell$ states, each an $\ell-$bit Haar random state. Let $\vec{\alpha} = (\vec{\alpha_1}, \cdots, \vec{\alpha_m})$ where each $\vec{\alpha_i}\in \{0, x, y, z\}^{\ell}$. It follows that 
$$
\langle C_{\vec{\alpha}}^2\rangle = \langle C_{\vec{\alpha_1}}^2\rangle \langle C_{\vec{\alpha_2}}^2\rangle\cdots \langle C_{\vec{\alpha_m}}^2\rangle
$$
Moreover, if the pauli weight $k(\vec{\alpha})=k$, at-most $m-k$ of $\langle C_{\vec{\alpha_i}}^2\rangle$s are $1/4^{\ell}$. Therefore, 
$$
\langle C_{\vec{\alpha}}^2\rangle  \leq \frac{1}{4^{\ell (m-k)}}\frac{1}{4^{\ell k }(2^{\ell}+1)^k} = \frac{1}{4^N (2^{\ell}+1)^k}
$$
Thus, in general, the weight $w \lessim \frac{1}{(2^{\ell}+1)^k}$ falls down exponentially in $k$(Fig.~\ref{FigS2}), as seen in Figure $2$d of the main text. 

\section{Short time dynamics}
In this section, we show a few basic results on the short time dynamics of $||\error_0 - \langle \error(t) \rangle_{\mu} ||$. We first show an upper bound on its magnitude 
$$
||\error_0 - \langle \error(t) \rangle_{\mu} || \leq ||\error_0|| +\left|\left|\int d\mu(\error) \error(t) \right|\right|
$$
The time evolution of a particular $\error$ is unitary and therefore, $||\error(t)||=||\error||$. This, along with the triangle inequality gives 
$$
||\error_0 - \langle \error(t) \rangle_{\mu} || \leq ||\error_0|| +\int d\mu(\error) ||\error || \leq ||\error_0|| +\sqrt{||\error_0||^2 +\sigma_0^2}
$$
The last inequality follows from the identity $\int |X| \leq \sqrt{\int |X|^2}$.

\section{Measuring other symmetries of the Hamiltonian}
In this section, we discuss the case of using other operators that commute with $H_{target}$. If $\hat{O}$ any operator that commutes with $H_{target}$, we will show an expression similar to Eq. ~($6$) of main text, corresponding to the time evolution of the operator under $H_{target}+\error$. One can define $\operator$ as the non-commuting part of  $\hat{O}$ w.r.t $H_{target} + \error$. Note that $\operator=0$ when $\error=0$. It follows that 
\begin{equation}\label{Operatro_dynamics}
    \hat{O}(t) - \hat{O}(0) = \operator- \operator(t)
\end{equation}

If $H_{target}$ doesn't have any degeneracies (or if $\hat{O}$ and $H_{target}$ have the same structure of degenerate blocks), we can show that $||\operator|| =O(||\error||)$. Indeed, if $\{\ket{\psi_n}\}$ are the common eigenstates of $H_{target}$ and $\hat{O}$ and $\{\ket{\tilde{\psi}_n}\}$ are the eigenstates of $H_{target}+\error$, in the absence of degeneracies in the eigenvalues of $H_{target}$, it follows that 
$$
\ket{\tilde{\psi}_n} = \ket{\psi_n} + O(||\error||)
$$
Therefore, the non-commuting part $\operator$ or $\hat{O}$ is ($m\neq n$)
\begin{equation}
    \operator_{mn} = \bra{\tilde{\psi}_m}\hat{O}\ket{\tilde{\psi}_n} = O(||\error||)
\end{equation}
Therefore, one can use Eq.~(\ref{Operatro_dynamics}) to estimate $||\error||$ by measuring $\hat{O}(t)-\hat{O}(0)$ in these cases. The magnitude of $\operator$ is much more complicated if $H_{target}$ has degenracies where $\hat{O}$ doesn't. For instance, the error can be magnified in this case. 

\subsection{Higher powers of $H_{target}$}
In this section, we consider the idea of measuring higher powers of $H_{target}$ in order to estimate $\error$. We show that while measuring $H_{target}^k$ more data-expensive, it is more sensitive to the error. In fact, the sensitivity increases exponentially in $k$.  
In section~\ref{op_therm}, we showed that the thermal value of any operator is its commuting part. Indeed,
$$
\lim_{T\rightarrow \infty}\frac{1}{T}\int_0^T H_{target}^k(t) - H_{target}^k(0)  = (H_{target}^k)_{com} - H_{target}^k
$$
We now use $(H_{target}+\error)^k_{com} = (H_{target}+\error)^k$. Thus, using what we showed in section~\ref{commuting_part}, 
$$
 (H_{target}+\error)^k_{com} = (H_{target})^k_{com} + (\error H_{target}^{k-1})_{com} + \cdots 
$$
Therefore, 
$$
(H_{target}^k)_{com} - H_{target}^k = (\error H_{target}^{k-1})_{com} + \cdots
$$

For higher values of $k$, the magnitude of the leading order is about $O(k ||H_{target}||^{k-1}||\error_{noncom}||)$ Therefore, the sensitivity increases exponentially in $k$. However, measuring the terms in $H_{target}^k$ would require much larger dataset -- also exponential in $k$. Moreover, unlike for the case of $k=1$, it is nontrivial to estimate the various components of $\error$ using measurements of $H_{target}^k$.

\section{Proof of theorem 1}
In this section, we will provide a detailed proof of theorem 1. The proof uses four lemmas, listed and proved below. We need to develop some background before that. The function $ \error(t)$ can be written as
\begin{equation}\label{adjoint}
\error(t) =e^{i t (\text{ad}_{H_{target}} + \text{ad}_{\error}) } \error
\end{equation}
 
Here, $\text{ad}_X A = [X, A]$ is the \textit{adjoint}, i.e., commutator super-operator, acting on the vector space of operators. We begin a few elementary properties of the adjoint.

\subsection{Properties of the adjoint of an operator}
The adjoint map, $X\mapsto \text{ad}_X$ that takes an operator to its adjoint has a $1$D kernel --- the space spanned by identity. That is, $\text{ad}_{\mathbbm{1}}=0$. Therefore, $\text{ad}_X = \text{ad}_{X-\lambda \mathbbm{1}}$ for any scalar $\lambda$. In other words, we may assume that $X$ to be traceless without loss of generality. We show the following properties, when $\text{Tr}(X)=0$:
\begin{itemize}
\item[i.] $\text{ad}_{\lambda_1 X_1 + \lambda_2 X_2}= \lambda_1 \text{ad}_{X_1}+\lambda_2 \text{ad}_{X_2}$ for scalars $\lambda_i$.
    \item[ii.] $\text{Tr}(X_1X_2^T) = 2d\text{Tr}(\text{ad}_{X_1}\text{ad}_{X_2}^T)$.
    \item [iii.] $\sigma_{\max}(X)< \sigma_{\max}(\text{ad}_X)\leq 2 \sigma_{\max}(X)$
\end{itemize}

i. and ii. together imply that the adjoint is a homeomorphism of the space of all traceless matrices. i. follows from the definition of the adjoint. To show ii., we begin by showing that $\text{ad}_X^T = \text{ad}_{X^T}$. If $E_{ij}=\ket{i}\bra{j}$ with $i, j =1, 2, \cdots, d$ are the basis elements of the space of $d\times d$ matrices, it follows that 
$$
(\text{ad}_{X})_{ij, kl} = \text{Tr}(E_{ji}[X, E_{kl}] ) = \delta_{lj}X_{ik}-\delta_{ik}X_{lj}
$$
Thus, 
$$
(\text{ad}_{X})^T_{ij, kl} = (\text{ad}_{X})_{kl, ij} = \delta_{lj}X_{ki}-\delta_{ik}X_{jl} = \delta_{lj}X^T_{ik}-\delta_{ik}X^T_{lj}
$$
It therefore follows that $\text{ad}_X^T = \text{ad}_{X^T}$. Now let's consider $\text{Tr}(\text{ad}_X\text{ad}_Y)$
\begin{equation}
\begin{split}
    \text{Tr}(\text{ad}_X\text{ad}_Y) =& \sum_{ij} \text{Tr}(E_{ji} [X, [Y, E_{ij}]])\\
    =& 2d \text{Tr}(XY)-2\text{Tr}(X)\text{Tr}(Y)
\end{split}
\end{equation}

ii. now follows. To prove iii., note that if the eigenvalues of $X$ are  $\{\lambda_i\}$, the eigenvalues of $\text{ad}_X$ are $\lambda_i-\lambda_j$. Tracelessness of $X$ implies that there is altleast one positive and one negative eigenvalue. Therefore, the maximum eigenvalue $\lambda_{\max}>0$ and the minimum eigenvalue is $\lambda_{\min}<0$. Moreover, $\sigma_{\max}(X) = \max\{\lambda_{\max}, -\lambda_{\min}\}$. It follows that $\sigma_{\max}(\text{ad}_X)=\lambda_{\max}-\lambda_{\min}$. Thus, iii. follows.  

\subsection{Bounds on the averages of oscillatory functions}
We will begin with an intuition for why the theorem should hold. Let us consider the function $f(x) = e^{ix} $. If $\epsilon$ is a gaussian random number with $\langle \epsilon \rangle =0$ and $\langle \epsilon^2 \rangle =\sigma_0^2$, we consider the mean $\langle \epsilon f(\epsilon t)\rangle$. Although the leading order of $f(\epsilon t) \epsilon$ is $\epsilon^2$,  $|\langle \epsilon f(\epsilon t)\rangle|  =  \sigma_0^2 t e^{-t^2\sigma_0^2/2}$, which reaches it's maximum at $t=1/\sigma_0$, with a value $\sigma_0e^{-1/2}$. We will show a stronger version of this statement and use it to prove theorem 1. 

\noindent \textbf{Lemma 2:} For a scalar valued function $f(x)$ with $f(0)=0$ and $|f'(x)|\geq b$ for $x\in (0, a)$ and $\epsilon \sim \mathcal{N}(0, \sigma_0)$, 
$$
\max_t |\langle \epsilon f(\epsilon t)\rangle| \geq \frac{ab}{2} \sigma_0
$$
\noindent \textbf{Proof: } Using the mean value theorem, 
$$
f(x) = x f'(\xi)
$$
for $\xi < x$. Thus, setting $t=a/(2\sigma_0)$, it follows that 
$$
\epsilon f(\epsilon a/\sigma_0)  \geq \frac{ab\epsilon^2}{\sigma_0}
$$
when $|\epsilon|\leq \sigma_0 $. Following the weight of the distribution over $|\epsilon|\geq \sigma_0$ if less than $1/2$, the result follows $\blacksquare$.

Clearly, this will be useful to prove some general statements regarding averages of oscillatory functions. However, there are two problems: (i) the function $\error(t)$, while being oscillatory, takes a matrix input $\error$ instead of a scalar $\epsilon$ and (ii), this function is matrix-valued unlike $f$ above which is scalar valued.  Below, we undertake the task of generalizing lemma 1, to matrix-valued functions which take matrix arguments. We start with a small step by considering functions that take vector arguments. That is, $f(\textbf{x})$, where $\textbf{x}\in \mathbbm{R}^k$ for some integer $k$. The first derivative is the gradient vector $f'(\mathbf{x})$ and the second derivative is the Hessian matrix $f''(\textbf{x})$ given by
\begin{equation}
    \begin{split}
        [f'(\textbf{x})]_i = \frac{\partial f}{\partial x_i}\\
        [f''(\textbf{x})]_{ij} = \frac{\partial^2 f}{\partial x_i \partial x_j}\\
    \end{split}
\end{equation}
Correspondingly, the error term $\bm{\epsilon}$ is also a vector with $k$ components and we set $\langle \bm{\epsilon}\rangle=0$ and the second moment $\Sigma$, defined as
$$
\Sigma_{ij} = \langle {\epsilon_i \epsilon_j}\rangle
$$
The covariance matrix $\Sigma$ determines the second moments. For this case, we prove the following equivalent of lemma 2. 

\noindent \textbf{Lemma 3:} For a scalar valued function $f(\textbf{x})$ with a vector argument $\mathbf{x}$ of dimension $k$, with $f(\mathbf{0})=0$ and $||f'(\mathbf{x})||=a$ and $\sigma_{\max}(f''(\textbf{x})) \leq c$ and $\bm{\epsilon} \sim \mu(\bm{\epsilon})$ with $\mu$ being light-tailed, 
$$
\max_t ||\langle \bm{\epsilon} f(\bm{\epsilon} t)\rangle|| \geq \frac{a^2 \sigma^2_{\min}(\Sigma)}{4 b k \sigma_{\text{avg}}(\Sigma)}
$$
where $\sigma^2_{\min}(\Sigma)$ is the minimal singular value of the covariance matrix $\Sigma$ of $\mu$ and $\sigma^2_{\text{avg}} =\frac{1}{k}\text{Tr}(\Sigma)$ is the average singular value. 

\noindent \textbf{Proof: } We have to use the mean value theorem, as before. It follows that
$$
f(t\bm{\epsilon})=t \bm{\epsilon}\cdot f'(s\bm{\epsilon})
$$
for some $s<t$. The trouble here is that the function $f'(s\bm{\epsilon})$ is a vector, and it varies with $\bm{\epsilon}$. Therefore, it enters the averaging over the measure, unlike in the previous case. In fact, $f'(s\bm{\epsilon})$ typically spans over the ellipsoid defined by $\Sigma$. So we need to understand this span before we proceed. We define:

$$
S=\{f'(\textbf{x}):\ ||\textbf{x}||\leq r\}
$$
This is the span of the vectors $f'(\textbf{x})$ when $\textbf{x}$ is in a sphere of radius $r$. We consider such a sphere because, the light-tailed measure $\mu$ will be dominant over a sphere centred at the origin with some radius to be determined using the second moments $\Sigma$.  In other words,  for each $\delta>0$there exists an $r=O(\sqrt{\text{Tr}(\Sigma)})$, $\int_{||\mathbf{x}||\leq r} d\mu(\textbf{x}) \geq 1-\delta $. We will again use the mean value theorem to understand $S$. 
$$
f'(\textbf{x}) = f'(\textbf{0})+f''(s \textbf{x})\textbf{x}
$$
for some $s<1$. We have assumed a bound $\sigma_{\max}(f''(\textbf{x}))\leq c$. Using this,
$$
||f'(\textbf{x})-f'(\textbf{0})||\leq c ||\textbf{x}||\leq c r
$$
Thus, $S$ is contained inside a sphere centered at $f'(\textbf{0})$, with radius $cr$. Given that a sphere is convex, a weighted average of vectors in $S$ will also be contained in the same sphere. However, it will be clear soon that we want \textit{more} than a weighted average --- we want the image of $S$ under averaging over a positive operator valued measure (POVM). That is,  we consider
$$
\tilde{S}=\left\{\sum_i \hat{\mu}_i \textbf{v}_i:\ \textbf{v}_i\in S \right\}
$$
where $\{\hat{\mu}_i\}$ is a set of $k\times k$ positive semi-definite (PSD) matrices, i.e., $\hat{\mu}_i \succeq 0$ and $\sum_i \hat{\mu}_i = \mathbbm{1}$. This is similar to taking a convex hull, but more general. 
This set lies inside a sphere centered also at $f'(\textbf{0})$, but with a radius $\sqrt{k}cr$. To see this, consider
$$
||\sum_i \hat{\mu}_i \textbf{v}_i-f'(0)||= ||\sum_i \hat{\mu}_i (\textbf{v}_i-f'(0))||
$$
The vector $\textbf{v}_i-f'(0)$ can be considered as any vector with a length bounded by $cr$. We use $\mathbf{u}_i$ to represent this. Moreover, the PSD matrices $\hat{\mu}_i$ can be written as a sum of rank-1 projectors and therefore a POVM can be considered as a measure on the unit sphere in $k-$ dimensional space.  In other words,  the above expression can be written as
$$
\sum_i \hat{\mu}_i  \textbf{u}_i = \int d\tilde{\mu}(\textbf{x}) \textbf{x}\textbf{x}^T \textbf{u}(\textbf{x}) 
$$
Here, $\textbf{x} \textbf{x}^T$ is a $1D$ projector, i.e., $||\textbf{x}||=1$, coming from the eigenvectors of $\hat{\mu}_i$ and the eigenvalues are absorbed into the measure $\tilde{\mu}(\textbf{x}) $. Note that this measure is a 2-design, because
\begin{equation}\label{2Design}
 \int d\tilde{\mu}(\textbf{x}) \textbf{x}\textbf{x}^T = \mathbbm{1}
\end{equation}
In particular, it also follows that for each component $x_i$ of the vector $\textbf{x}$, 
$$
\langle x_i^2 \rangle_{\tilde{\mu}}=1
$$
and, $\langle ||\textbf{x}||^2 \rangle_{\tilde{\mu}}=k$. Also note that the R.H.S of Eq.(\ref{2Design}) is invariant under rotations.  Therefore, $\langle x_i^2 \rangle_{\tilde{\mu}}=1$ holds in \textit{any} basis.  We are now to estimate the magnitude of the vector $\textbf{w}= \int d\tilde{\mu}(\textbf{x}) \textbf{x}\textbf{x}^T \textbf{u}(\textbf{x})$ with $||\textbf{u}(\textbf{x})||\leq cr$.  Without loss of generality, we may assume that the vector $\textbf{w}$ is parallel to the first of the basis elements, given that  Eq.(\ref{2Design}) is invariant under rotations.  Thus, 
$$
 ||\textbf{w}||=|\langle  x_1 \textbf{x}\cdot \textbf{u}(\textbf{x})\rangle| \leq cr\langle  |x_1|\  ||\textbf{x}||\rangle 
$$
We will now use Cauchy-Schwartz inequality to show that $\langle  |x_1| \  ||\textbf{x}||\rangle \leq \sqrt{\langle  x_1^2\rangle \langle   ||\textbf{x}||^2\rangle } = \sqrt{k}$.  

Thus, if we choose the original radius $r$, which depends on the moments $\Sigma$ of the measure $\mu$ to be small enough, we can show that
\begin{equation}
    ||\textbf{v}||\geq a -\sqrt{k} c r, \ \textbf{v}\in \tilde{S}
\end{equation}
Recall that $||f'(\textbf{0})||=a$. In particular, we can exclude $\textbf{0}$ from $\tilde{S}$ if we choose for example,
$$
r = \frac{a}{2\sqrt{k}c}
$$
The vectors in $\tilde{S}$ would satisfy $||\textbf{v}||\geq a/2$. This will be useful soon. The bottom line is, $\tilde{S}$ is contained in a sphere centered at $f'(\textbf{0})$, and a radius $\sqrt{k}cr$. We are now ready to get back to the main expression:
$$
\langle \bm{\epsilon} f(t\bm{\epsilon})\rangle = t \langle \bm{\epsilon} f'(s_{\bm{\epsilon}}\bm{\epsilon})\cdot \bm{\epsilon}\rangle 
$$
for some $s_{\bm{\epsilon}}<1$. This is non-trivial to evaluate. However, let us define 
$$
\hat{\mu}_{\bm{\epsilon}} = \Sigma^{-1} \bm{\epsilon} \bm{\epsilon}^T
$$
Note that $\hat{\mu}_{\bm{\epsilon}}  \succ 0$ and $\int d\mu(\epsilon) \hat{\mu}_{\bm{\epsilon}} =\mathbbm{1}$. In other words, this is a POVM. Therefore, returning to the previous equation, 
$$
\int \hat{\mu}_{\bm{\epsilon}} f'(s_{\bm{\epsilon}}) = \mathbf{v}\in \tilde{S}
$$
And,
$$
\langle \bm{\epsilon} f(t\bm{\epsilon})\rangle = t \Sigma \textbf{v}
$$
Note that all this is true only if $r$ is chosen to include the bulk of the distribution $\mu$. In other words, we need 
$$
t= \frac{r}{\sqrt{\text{Tr}(\Sigma)}} = \frac{a}{2\sqrt{k} c \sqrt{\text{Tr}(\Sigma)}}
$$
so that the vector $\bm{\epsilon}t$ remains mostly within the sphere of radius $r$, while averaging over $\mu$. We now use the fact that $||\mathbf{v}||\leq a/2$ for this choice of $r$. Therefore, 
$$
\max_t ||\langle \bm{\epsilon} f(t\bm{\epsilon})\rangle||  \geq \frac{a^2}{4\sqrt{k} c \sqrt{\text{Tr}(\Sigma)}} \sigma^2_{\min}(\Sigma)
$$
This proves the lemma $\blacksquare$

We make a few comments before proceeding further. If the singular values of $\Sigma$ dont have a large variance and we can suppose that there is a typical value $\sigma_0^2$. This will result in
$$
\max_t ||\langle \bm{\epsilon} f(t\bm{\epsilon})\rangle|| \geq \frac{a^2  \sigma_0}{4k c}
$$
Also, $a\sim O(\sqrt{k})$, since $f'(\textbf{0})$ is a $k-$ dimensional vector. Therefore, we recover the scaling we had in the previous lemma. 

This works well for scalar functions. We need similar results for vector valued functions. We begin with generalizing lemma 1a for vector valued functions. The mean-value theorem, which plays an important role in this proof reads different for vector valued functions. If $\textbf{f}(x)$ is a vector valued function, the vector MVT~\cite{mcleod_1965} reads,
\begin{equation}\label{MVT}
    \textbf{f}(x)-\textbf{f}(x_0) = (x-x_0)\sum_{i=1}^n \mu_i \textbf{f}'(x_i)
\end{equation}
where, $n$ is dimension of the vectors $f(x)$, $x_0 < x_i < x$ and $\mu_i \geq 0 , \sum_i \mu_i =1$ are non-negative weights. Geometrically, the mean values of $f$ lie in the convex hull of its derivatives. Similarly, for vector valued functions with vector argument, 
\begin{equation}\label{VMVT}
    \textbf{f}(\textbf{x})-\textbf{f}(\textbf{x}_0) = \sum_i \mu_i \textbf{f}'(\textbf{x}_i)(\textbf{x}-\textbf{x}_0)
\end{equation}
Here, $\textbf{x}\in \mathbbm{R}^k$ and $\textbf{f}(\textbf{x})\in \mathbbm{R}^n$. $\textbf{x}_i = s_i \textbf{x}_0 + (1-s_i)\textbf{x}$  for some $0<s_i<1$. Note that $\textbf{f}'(\textbf{x})$ is the Jacobian matrix, defined by 
$$
[\textbf{f}'(\textbf{x})]_{ij} = \frac{\partial f_i(\textbf{x})}{\partial x_j}, \ i =1, \cdots, n \ \text{ and } j=1, \cdots, k
$$
The second derivative is a rank-3 tensor, defined by 
$$
[\textbf{f}''(\textbf{x})]_{ijl}=\frac{\partial^2 f_i(\textbf{x})}{\partial x_j\partial x_l}, \ i =1, \cdots, n \ \text{ and } j, l=1, \cdots, k
$$

\noindent \textbf{Lemma 4:} For a vector valued function $\textbf{f}(\textbf{x})$ with a vector argument $\mathbf{x}$ of dimension $k$, with $\textbf{f}(\mathbf{0})=\textbf{0}$ and $||\textbf{f}'(\mathbf{x})||=a$ and $\sigma_{\max}(\textbf{f}''(\textbf{x})) \leq c$ 
$$
\max_t || \langle \bm{\epsilon} \textbf{f}(\bm{\epsilon} t)\rangle || \geq \frac{a^2 \sigma^2_{\min}(\Sigma)}{4 b k \sigma_{\text{avg}}(\Sigma)}
$$
where, $\bm{\epsilon} \textbf{f}(\bm{\epsilon} t)$ is a \textit{free} product, i.e., outer product between vectors.  This would be a $k\times n$ matrix. 

\noindent \textbf{Proof: } We again start with the mean value theorem
$$
\textbf{f}(t\bm{\epsilon})=t  \sum_j \mu_j \textbf{f}'(s_j\bm{\epsilon}) \bm{\epsilon} = t J_{\bm{\epsilon}} \bm{\epsilon}
$$
Here, $J_{\bm{\epsilon}}= \mu_j \textbf{f}'(s_j\bm{\epsilon}) $. We define a convex set, 
$$
S= \text{Convex.Hull}\{\textbf{f}'(\textbf{x}): \ ||\textbf{x}|| \leq r\}
$$
As before, we will try to understand the set $S$. Again, using Eq.~(\ref{VMVT}), 
$$
\textbf{f}'(\textbf{x}) = \textbf{f}'(\textbf{0}) + \sum_{j=1}^{nk} \mu_j \textbf{f}''(\textbf{x}_j) \textbf{x}
$$
It follows that 
$$
||\textbf{f}'(\textbf{x}) - \textbf{f}'(\textbf{0}) || \leq c ||\textbf{x}|| = cr
$$
It is important to note that $c$ is a bound on the maximum singular value of $\textbf{f}''(\textbf{x})$, treated as a linear map from a $k-$ dimensional space to a $nk$ dimensional space.  The rest of the proof is very similar to the previous lemma.

Note that, following Eq.~(\ref{adjoint}), $\error (t) = e^{i t (\text{ad}_{H_{target}} + \text{ad}_{\error}) } \error$, with $\error =\sum \epsilon_{\vec{\alpha}}P_{\vec{\alpha}}$. So, we try to apply the theorem to $\textbf{f}(\error) = e^{it\text{ad}_{H_{target}} + i t \text{ad}_{\error} }$. We need a lemma to show the lower bound on the derivatives of such functions. $i\text{ad}_X$ is anti-hermitian for hermitian $X$. Moreover, $\error = \sum \epsilon_{\vec{\alpha}}P_{\vec{\alpha}}$. We therefore show the following lemma.

\noindent \textbf{Lemma 5:} If $A, B_1, \cdots, B_k$ are anti-hermitian matrices and $\textbf{f}(\textbf{x}) = e^{A+x_1B_1+\cdots+x_k B_k}$ is a unitary matrix-valued function, $||\textbf{f}''(\textbf{x})||\leq \sum ||B_i||^2$. Moreover, $||\textbf{f}'(\textbf{0})|| = O(\sqrt{\sum_i ||B_i||^2})$.

\noindent \textbf{Proof:} Differentiating the function $\textbf{f}(\textbf{x})$ is quite non-trivial. We will show some elementary results regarding a related function:
$$
\textbf{g}(x) = e^{A + Bx}
$$
for anti-hermitian $A$ and $B$. We will use Schrieffer-Wolff transformation for small $x$ to show that $||g'(0)|| = O(||B||)$. Under the Schrieffer-Wolff transformation~\cite{BRAVYI20112793}, assuming $x$ is small,  $\exists$ an anti-hermitian operator $S$, such that 
$$
e^{-xS}Ae^{xS} =  A + x B_{noncom} + O(x^2)
$$
Here, $B_{noncom}$ is the non-commuting part of $B$ w.r.t $A$. Explicitly, $S$ is a solution to $[A, S]=B_{noncom} \implies \text{ad}_A S =B$.We now define 
\begin{equation}
\begin{split}
\textbf{h}(x)&=e^{e^{-xS}(A+ xB_{com})e^{xS}}\\
&=e^{-xS}e^{A+xB_{com}}e^{xS} = \textbf{g}(x)+O(x^2)
\end{split}
\end{equation}

Here, $B_{com}$ is the commuting part of $B$ w.r.t $A$. It follows that $g'(0)=f'(0)$. Thus, 
$$
\textbf{g}'(0) = \textbf{h}'(0) = B_{com}e^A + [e^A, S] 
$$
Note that $e^A$ is a unitary. Therefore, estimating $||\textbf{g}'(0)||$ boils down to estimating $||S||$. The equation $\text{ad}_A S =B_{noncom}$ defines $S$, and potentially, the $||S||$ could scale as $||B||/\sigma_{max}(\text{ad}_A)$. Noting that the singular values of $\text{ad}_A$ are $|\lambda_i-\lambda_j|$ where $\{\lambda_i \}$ are the eigenvalues of $A$, this can be potentially very large. However, in our application, $A=it H_{target}$. Therefore, we will use the fact that the typical value of the gaps $|\lambda_i-\lambda_j|$ is $O(1)$, for a random matrix~\cite{santos2017nonequilibrium, haake2010quantum}.  It now follows that $||g'(0)|| = O(||B||)$. Thus, it follows that $||\textbf{f}'(\textbf{0})|| = O(\sqrt{\sum_i ||B_i||^2})$.

Let us define the function $f_N(x)$ for integers $N$:
$$
\textbf{f}_N(\textbf{x}) = \left( \mathbbm{1} +\frac{A+x_iB_i}{N}\right)^N
$$
Let us also assume that $N$ is large enough so that $\textbf{f}_N(\textbf{x})^{-1}$ exists for the given $N$. It follows that 
\begin{equation}
\begin{split}
    \frac{\partial^2 \textbf{f}_N(\textbf{x})}{\partial x_i \partial x_j} &= \frac{1}{N^2}\sum_{p,q}  \left( \mathbbm{1} +\frac{A+x_iB_i}{N}\right)^{p-1} B_i\\
   & \times \left( \mathbbm{1} +\frac{A+x_i B_i}{N}\right)^{q-1} B_j \left( \mathbbm{1} +\frac{A+x_iB_i}{N}\right)^{N-p-q}
\end{split}
\end{equation}
Using the triangle inequality and  setting $N\rightarrow \infty$, it follows that $\sigma_{\max}(\textbf{f}''(\textbf{x}))\leq \sqrt{\sum_i||B_i||^2} \blacksquare$. 

If $||B_i||=O(||B||) \forall i$, the results reduce to $||\textbf{f}'(\textbf{0})||=\sqrt{k}O(||B||)$ and $||\textbf{f}''(\textbf{x})||\leq k||B||^2$.

\subsection{Proof of theorem 1}
We can now use lemma 4 \& 5 to prove theorem 1. Let $\error = \sum_{\vec{\alpha}}\epsilon_{\vec{\alpha}} P_{\vec{\alpha}}$.  We may write $\error(t)$ as
$$
\error(t) = e^{-it \text{ad}_{H_{target}} -it \sum \epsilon_{\vec{\alpha}} \text{ad}_{P_{\vec{\alpha}}}}\sum_{\vec{\alpha}}\epsilon_{\vec{\alpha}} P_{\vec{\alpha}}
$$
We may treat the matrices $P_{\vec{\alpha}}$ as vectors. The product above is a matrix-product. It is a contraction of the outer product and therefore, we may apply lemma $4$ and $5$. It follows that 
$$
\max_t|| \langle e^{-it \text{ad}_{H_{target}} -it \epsilon \text{ad}_P}\epsilon\rangle || \geq O(\sigma_0)
$$

\section{Details of the Protocols}
\textbf{Protocol 1: } The theory developed in the main text uses the state $\rho_{\vec{\alpha}}$. While this can be produced experimentally, we construct a slightly different state for the protocol in order to minimize the number of sampling rounds.  For $\vec{\alpha}$ with $k(\vec{\alpha})=n$, let $\vec{\alpha}_1, \vec{\alpha}_2, \cdots, \vec{\alpha}_n$  be the weight-1 pauli indices such that $P_{\vec{\alpha}}=P_{\vec{\alpha}_1} P_{\vec{\alpha}_2}\cdots P_{\vec{\alpha}_n}$. For instance, if $\vec{\alpha}=xy000$, then it's weight is $n=2$ and $\vec{\alpha_1}=x0000$ and $\vec{\alpha_2}=0y000$. We now define the state  $\tilde{\rho}_{\vec{\alpha}} =  \frac{1}{2^N}(\mathbbm{1}^{\otimes N}+P_{\vec{\alpha}_1})(\mathbbm{1}^{\otimes N}+P_{\vec{\alpha}_2})\cdots (\mathbbm{1}^{\otimes N}+P_{\vec{\alpha}_N})$. In the above example, $\tilde{\rho}_{\vec{\alpha}}=\frac{1}{2^5}(\mathbbm{1}+\sigma_x)\otimes (\mathbbm{1}+\sigma_y)\otimes \mathbbm{1}^{\otimes 3} $. This state can be prepared with a fidelity of $\sim f^n\left(1-\frac{1}{\sqrt{\nu}\sqrt{1+2^{\ell}}}\right)$ using single qubit gates of fidelity $f$ and $\nu$ sampling rounds and $\ell-$bit entangled states, following the protocol presented in ref.~\cite{madhusudhana2023benchmarking1}. This state also contains $P_{\vec{\alpha}}$ and lower-weight Paulis. It takes fewer samples to produce this state than $\rho_{\vec{\alpha}}$. Let $H_{target}= H_1 + \cdots + H_r$, where each $H_j$ is a sum of mutually commuting Paulis. For instance, in example 1, Eq.~(\ref{tweezer}), $H_{target}=H_1+H_2$ with $H_1=\Omega \sum_j \sigma_{x, j}$ and $H_2 = \sum_j \Delta \sigma_{z, j} +\sum_j V \sigma_{z, j}\sigma_{z, j+1} $. One needs separate measurement settings to measure each $H_j$.  Protocol 1 is based on measuring each $H_j$ after time evolution of the states $\tilde{\rho}_{\vec{\alpha}}$ and using the result to extract the coefficient $\langle \epsilon_{\vec{\alpha}}\rangle_{\mu}$. We present this protocol in Fig.~\ref{protocols}, left panel. 

In this protocol, for each $n$,  we need $3^n \binom{N}{n} r $ measurement settings and $\nu$ sampling rounds. From these measurement results, one can extract $\langle \epsilon_{\vec{\alpha}}\rangle_{\mu}$ for $k(\vec{\alpha})\leq m$.  The number of measurement settings scales polynomially in $N$, which could still be nontrivial to implement. Moreover, it also needs single qubit addressing, which is not available in all experiments.

\subsection{Protocol for symmetric variables}

\begin{figure}
    \centering
    \fbox{\begin{minipage}[r]{0.45\textwidth}
    \textbf{Protocol 3 (Symmetric states)}
    \begin{itemize}
    \item Pick $p$ random pure states, represented by points $(x_i, y_i, z_i)$ on the Bloch sphere for $i=1, 2, \cdots, p$
    \item Repeat for $i\in \{1, 2, \cdots, p\}$:
    \begin{itemize}
        \item Repeat for  $j \in \{1, 2, \cdots, r\}$ :
        \begin{itemize}
            \item[0] (Repeat for $\nu$ samples)
            \item[1] Prepare all the $N$ qubits in the state represented by $(x_i, y_i, z_i)$ on the Bloch sphere.
            \item[2] Evolve this initial state under the experimentally implemented target Hamiltonian $H_{target}$.
            \item[3] Apply the single qubit gate on each of the $N$ qubits appropriate to measure $H_j$. 
            \item[4] Measure in the $z$ basis.
        \end{itemize}
    \end{itemize}
\end{itemize}
\vspace{5pt}
  \end{minipage}}
    \caption{\textbf{Protocol:} shows a polynomial interpolation protocol, which can be implemented without single-qubit addressing. It can be used to extract the symmetric, i.e., $S_N-$invariant components $\langle \epsilon_{k_x, k_y, k_z}\rangle_{\mu}$ of the error (see text). }
    \label{protocols}
\end{figure}
Below, we develop a simpler protocol, that doesn't use single-site addressing. That is, we will only use initial states of the form $\rho^{\otimes N}$. However, this protocol can only measure the error components that are invariant under qubit permutations. 

A single qubit pure state is represented by a point  on the Bloch sphere, represented by three real coordinates $(x, y, z)$ with  $x^2 + y^2 + z^2=1$. It can be written as $\rho =\frac{1}{2}(\mathbbm{1}+x\sigma_x + y\sigma_y + z\sigma_z)$. If all the $N$ qubits are prepared in the same state, the $N-$qubit state expands as 
\begin{equation}
    \rho^{\otimes N} = \frac{1}{2^N}\sum_{\vec{\alpha}\in \{0,x,y,z\}^N }x^{k_x(\vec{\alpha})}y^{k_y(\vec{\alpha})}z^{k_z(\vec{\alpha})} P_{\vec{\alpha}}
\end{equation}
where $k_i(\vec{\alpha})$ is the number of $i$'s in $\vec{\alpha}$. Note that this is a polynomial in $x, y$ and $z$ and it collects together all the $P_{\vec{\alpha}}$ that are equivalent under the action of the permutation group $S_N$, forming invariant terms. We define these terms as:
$$
Q_{k_x, k_y, k_z} = \sum_{k_i(\vec{\alpha})=k_i}P_{\vec{\alpha}}
$$
These $Q$ operators are sums of paulis and they are invariant under $S_N$. Indeed,
$$
\rho^{\otimes N} = \frac{1}{2^N}\sum_{k_x, k_y, k_z} x^{k_x}y^{k_y}z^{k_z} Q_{k_x, k_y, k_z}
$$
This state, therefore, lives completely in the invariant subspace under the permutation group $S_N$. Measurements using this state can't be used to determine the variation of the error over the qubits. However, we can always extract the $S_N-$invariant parts of the error $\error_0$. It follows that, if we measure the dynamics of the Hamiltonian using this initial state 
\begin{equation}
\begin{split}
    \text{Tr}[\rho^{\otimes N} \error_0 ] =& \sum  x^{k_x}y^{k_y}z^{k_z} \langle\epsilon_{k_x, k_y, k_z}\rangle_{\mu} \text{ where, }\\
    \langle\epsilon_{k_x, k_y, k_z}\rangle_{\mu}=& \sum_{k_i(\vec{\alpha})=k_i} \langle\epsilon_{\vec{\alpha}}\rangle_{\mu}
\end{split}
\end{equation}
The coefficients $\langle\epsilon_{k_x, k_y, k_z}\rangle_{\mu}$, which represent the $S_N-$invariant parts of the error, can be estimated by using a polynomial interpolation of the Hamiltonian dynamics measured for various values of $x, y$ and $z$. This forms our second protocol (Fig.~\ref{protocols}).  This protocol needs $p r$ measurement settings and $\nu$ sample rounds. Typically, $r=O(1)$ and $p$ depends on $m$ as $p = O(3^m)$. One can develop variants of this protocol by focussing on the powers of $z$, for instance, whenever experimentally relevant. \\

\section{Main text figures}
In this section, we provide more details on how the data in the figures of main text were produced. 

\subsection{Main text figure 2}

\begin{figure}
    \centering
    \includegraphics{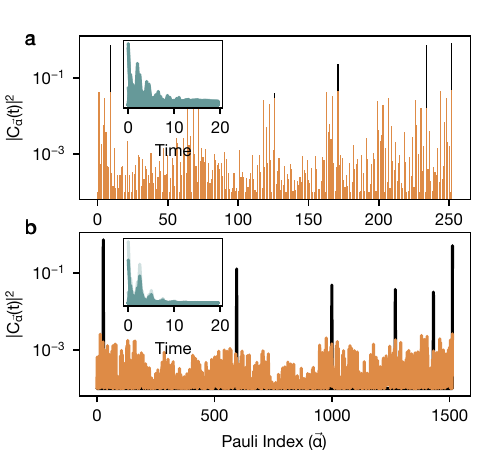}
    \caption{\textbf{Operator thermalization:} \textbf{a} shows $|C_{\vec{\alpha}}(t)|^2$ for $t=0$ (black) and $t=20$ (orange) for $k(\vec{\alpha})=2$ and $N=8$. \textbf{b} shows the same for $k(\vec{\alpha})=3$.}
    \label{FigS1}
\end{figure}
In figure $2$a, we used a Hamiltonian
\begin{equation}\label{Hamiltonian}
\begin{split}
    H_{target} & = \Omega \sum_j \sigma_{x, j} + \Delta \sum_j \sigma_{z, j} + V\sum_{j} \sigma_{j, z}\sigma_{j+1, z}, \\
    & \Omega=2,\  \Delta=1 \text{ and } V=1
    \end{split}
\end{equation}
We used $N=8$ qubits and a fixed error given by
$$
\error = \sum_j \epsilon_j \sigma_{z, j}
$$
where $\epsilon_j$ are a randomly chosen, fixed set of numbers. We used an initial state $\rho = \frac{1}{2^N} (\mathbbm{1}-\sigma_z)^{\otimes N}$. i.e., all qubits in the state $\ket{1}$. We evolve $\error$ under $H_{target}+\error$ in time and obtain the coefficients $C_{\vec{\alpha}}(t)=\frac{1}{2^N}\text{Tr}[\error(t) P_{\vec{\alpha}}]$. The main panel shows $|C_{\vec{\alpha}}(0)|^2$ as the black bars, for the $24$ values of $\vec{\alpha}$  such that $k(\vec{\alpha})=1$. These coefficients are given by $C_{\vec{\alpha}}(0)=\epsilon_j$ for $\vec{\alpha} = (0, 0, \cdots, 0, z , 0, \cdots 0)$, where the $z$ is in the $j$-th position. The $\vec{\alpha}$s are ordered as $(x, 0, 0, \cdots, 0), (y, 0, 0, \cdots, 0), (z, 0, 0, \cdots, 0)$ $, (0, x, 0, \cdots, 0), \cdots, (0, 0, \cdots, 0, z)$ on the x-axis. The orange bars shows $|C_{\vec{\alpha}}(t)|^2$. One can also look at $C_{\vec{\alpha}}(t)$ for higher weight $\vec{\alpha}$, but the picture remains the same. The inset shows the time evolution of these coefficients, for all $24$ values of $\vec{\alpha}$. 

Similar numerical analysis for $k(\vec{\alpha})=2, 3$ is shown in Fig.~\ref{FigS1}a and Fig.~\ref{FigS1}b respectively. The error $\error$ in these datasets were of the form $\sum_j \epsilon_j \sigma_{z, j}\sigma_{z, j+1}$ and $\sum_j \epsilon_j \sigma_{z, j}\sigma_{z, j+1}\sigma_{z, j+2}$ repectively. 


In Figure $2$b, we choose three pauli operators $P_1 = \mathbbm{1}^{\otimes N/2}\otimes \sigma_z \otimes \mathbbm{1}^{\otimes N/2-1}$, $P_2 = \mathbbm{1}^{\otimes N/2-1}\otimes \sigma_z\otimes \sigma_z \otimes \mathbbm{1}^{\otimes N/2-1}$ and $P_3 = \mathbbm{1}^{\otimes N/2-1}\otimes \sigma_z\otimes \sigma_z\otimes \sigma_z\otimes \mathbbm{1}^{\otimes N/2-2}$ and compute the weight of their non-commuting parts, $w$ w.r.t the Hamiltonian in Eq.~(\ref{Hamiltonian}). We use $N=4, 6, 8, 10, 12, 13$. 

In Figure $2$c, we compute the weights of paulis $P_k = \mathbbm{1}^{\otimes (N-k)/2}\otimes \sigma_{\alpha_1}\otimes \cdots \otimes \sigma_{\alpha_k}\mathbbm{1}^{\otimes (N-k)/2}$, where $\alpha_1, \cdots, \alpha_k$ are randomly chosen from $\{x, y, z\}$. We choose $k=3, 4, 5, 6, 7,8 ,9 $ and $N=10, 11, 12$.

In Figure $2$d main panel, we compute $\text{Tr}[\rho(H_{target}(t)-H_{target}(0))]$ for $N=7$. The error was modelled as
\begin{equation}
    \error = \epsilon_{x} \sum_{i}\sigma_{x, i} + \epsilon_{y} \sum_{i}\sigma_{y, i} + \epsilon_{z} \sum_{i}\sigma_{z, i}
\end{equation}
Where $\epsilon_x, \epsilon_y$ and $\epsilon_z$ are random numbers chosen with a randomly chose, and fixed mean value and a standard deviation of $0.1$. We use the Hamiltonian in Eq.~(\ref{Hamiltonian}) and average the resulting time trace over $100$ samples. The state $\rho$ is $\frac{1}{2^N}(\mathbbm{1}-\sigma_z)^{\otimes N}$.

\subsection{Main text figure 4}
In the following subsection, the numerical details for the Fig. ~$4$ in the main text are discussed.  In Fig.~$4$a,  we simulate the time dynamics of $\langle H_{target}(t)-H_{target}(0)\rangle_{\mu}$ for Eq.~(\ref{tweezer}) for $N=6$ qubits and parameters $\Omega=2, V=1, \Delta=2$ upto $t=5$.  The error is modelled by $\error = \epsilon \sum_j \sigma_{z, j}$ with $\epsilon \sim \mathcal{N} (0, \sigma_0)$. We average over $100$ samples  of $\epsilon$.  After evolving $H_{target}$ in time, under the Heisenberg picture,  we compute the Schatten-2 norm of $\langle \error(t)-\error(0)\rangle_{\mu} = \langle H_{target}(0)-H_{target}(t)\rangle_{\mu}$, \textit{after} averaging.  Fig.~$4$b shows the components of the operator $\langle \error(t)\rangle_{\mu}$ in the pauli basis. The blue bars show the same for $t=0$.

The time evolution is discretized for numerical simulation by employing the Suzuki-Trotter approximation. We tested the evolution with several time steps and consequently found that the appropriate time step was $0.01$. Consider the single qubit density matrix $\rho_0$ as:
\begin{equation}
\rho_0 = \frac{1}{2}\begin{bmatrix}
    1 & 0 \\
    0 & 1
\end{bmatrix}
\end{equation}
Then the N qubit state can be written as $\rho_{0}^{\otimes N}$.  For all the results in Fig. 4 in the main text, the system size has been fixed to $N=4$. In the main text, Fig. 4(c), we present the temporal evolution of the Hamiltonian's expectation value in the presence of non-Markovian errors. The black curve in main text Fig. 4(a) delineates the time evolution of the target Hamiltonian, while the pink curve portrays the system's time evolution, accounting for errors resulting from fluctuations in the coefficients of the Hamiltonian ($\omega, \Delta$, and V). All the coefficients for the target Hamiltonian are set to $1$ with added fluctuations, which are drawn randomly from a Gaussian distribution comprising $10,000$ samples, each with a central value of $1$ and a standard deviation of $0.1$. The ultimate expectation value of the Hamiltonian, incorporating errors, is computed by performing a statistical average 
across all $10,000$ samples. If the system is allowed to evolve for an extended time scale, it will be revealed that the expectation value of the Hamiltonian, when affected by errors, would finally converge to the expectation value of the target Hamiltonian.  Fig. $4$c shows the maximum deviation of the time-evolved Hamiltonian with the target Hamiltonian for various system sizes. The inset within Fig.  $4$c presents the expectation value of the maximum difference between the target Hamiltonian and the time-evolved Hamiltonian as a function of the $\sigma_{0}$.


\end{document}